\documentclass[acmsmall, screen]{acmart}

\usepackage{soul}
\usepackage{makecell}
\usepackage{array}
\usepackage{threeparttable}
\usepackage{multirow}
\usepackage{ragged2e}
\usepackage{pgfplots}
\pgfplotsset{compat=1.16}
\usepackage{algorithm}
\usepackage[noend]{algpseudocode}
\makeatletter
\def\BState{\State\hskip-\ALG@thistlm}
\makeatother

\usepackage{mathtools}

\AtBeginDocument{%
  \providecommand\BibTeX{{%
    \normalfont B\kern-0.5em{\scshape i\kern-0.25em b}\kern-0.8em\TeX}}}

\setcopyright{licensedusgovmixed}
\acmPrice{}
\acmDOI{10.1145/3485531}
\acmYear{2021}
\copyrightyear{2021}
\acmSubmissionID{oopsla21main-p301-p}
\acmJournal{PACMPL}
\acmVolume{5}
\acmNumber{OOPSLA}
\acmArticle{154}
\acmMonth{10}

\begin{document}

\title[Not So Fast: Understanding and Mitigating Negative Impacts of Compiler Optimizations on Code Reuse...]{Not So Fast: Understanding and Mitigating Negative Impacts of Compiler Optimizations on Code Reuse Gadget Sets}

\author{Michael D. Brown}
\affiliation{%
  \department{School of Computer Science}
  \institution{Georgia Institute of Technology}
  \city{Atlanta}
  \state{GA}
  \country{USA}
}
\email{mbrown337@gatech.edu}

\author{Matthew Pruett}
\affiliation{%
  \department{School of Electrical \& Computer Engineering}
  \institution{Georgia Institute of Technology}
  \city{Atlanta}
  \state{GA}
  \country{USA}
}
\email{matthew.pruett@gtri.gatech.edu}

\author{Robert Bigelow}
\affiliation{%
  \department{School of Computer Science}
  \institution{Georgia Institute of Technology}
  \city{Atlanta}
  \state{GA}
  \country{USA}
}
\email{rbigelow3@gatech.edu}

\author{Girish Mururu}
\affiliation{%
  \department{School of Computer Science}
  \institution{Georgia Institute of Technology}
  \city{Atlanta}
  \state{GA}
  \country{USA}
}
\email{girishmururu@gatech.edu}

\author{Santosh Pande}
\affiliation{%
  \department{School of Computer Science}
  \institution{Georgia Institute of Technology}
  \city{Atlanta}
  \state{GA}
  \country{USA}
}
\email{santosh.pande@cc.gatech.edu}


\begin{abstract}
Despite extensive testing and correctness certification of their functional semantics, a number of compiler optimizations have been shown to violate security guarantees implemented in source code. While prior work has shed light on how such optimizations may introduce semantic security weaknesses into programs, there remains a significant knowledge gap concerning the impacts of compiler optimizations on non-semantic properties with security implications. In particular, little is currently known about how code generation and optimization decisions made by the compiler affect the availability and utility of reusable code segments called gadgets required for implementing code reuse attack methods such as return-oriented programming. 

In this paper, we bridge this gap through a study of the impacts of compiler optimization on code reuse gadget sets. We analyze and compare 1,187 variants of 20 different benchmark programs built with two production compilers (\verb|GCC| and \verb|Clang|) to determine how their optimization behaviors affect the code reuse gadget sets present in program variants with respect to both quantitative and qualitative metrics. Our study exposes an important and unexpected problem; compiler optimizations introduce new gadgets at a high rate and produce code containing gadget sets that are generally more useful to an attacker than those in unoptimized code. Using differential binary analysis, we identify several undesirable behaviors at the root of this phenomenon. In turn, we propose and evaluate several strategies to mitigate these behaviors. In particular, we show that post-production binary recompilation can effectively mitigate these behaviors with negligible performance impacts, resulting in optimized code with significantly smaller and less useful gadget sets.
\end{abstract}




\maketitle

\section{Introduction}

The design and implementation of code optimizations in production compilers is primarily concerned with the performance and correctness of the resulting optimized code. The security impacts of these design choices are difficult to determine and even more difficult to quantify, which is a natural consequence of the implementation-dependent nature of security vulnerabilities. Further, security issues caused by compiler optimizations are also difficult to detect during correctness certification because they may not be captured by the operational semantics model used in correctness proofs~\cite{dsilva}. 

As a result, a significant knowledge gap exists regarding the impact of code optimization on security that has been the subject of several research papers~\cite{deng2017securing, proy, lim, belleville, simon, deng2018securing, besson}. Of particular note is prior work by ~\citet{dsilva} that has shown that compiler optimizations can introduce semantic security weaknesses such as information leaks, elimination of security-relevant code, and side channels despite being formally proven correct. Additionally, our recent work \cite{brown} suggests that compiler optimizations may introduce a fourth, non-semantic, class of security weakness: increased availability and quality of code reuse attack (CRA) gadgets. 

Gadget-based CRAs obviate the need to inject malicious code by chaining together snippets of the vulnerable program's code called gadgets to implement an exploit. The diversity and utility of gadgets available for creating exploits depends on the compiler-generated code in a program's binary and its linked libraries. Gadget-based CRAs are particularly insidious because they circumvent code injection defenses and can achieve Turing-completeness~\cite{shacham, bletsch2011jump, checkoway, sadeghi}. Numerous gadget-based CRA defenses have been proposed, however their adoption remains low due to weaknesses against increasingly complex attack patterns and runtime overhead costs~\cite{van2017dynamics, muntean2019analyzing, 10.5555/2337159.2337171, carlini2014rop, carlini2015control, davi2014stitching, evans}. 

\subsection{Motivation}
Our prior work demonstrated that software debloating transformations introduce new code reuse gadgets into debloated binaries at a high rate, despite an overall reduction in the total number of gadgets \cite{brown}. Further, we showed that in many cases the introduced gadgets are more useful to an attacker than the gadgets that were removed, negatively impacting security. Interestingly, we identified differences in compiler optimization behavior on bloated versus debloated source code as one cause of gadget introduction. In contrast to debloating transformations that are localized in nature, compiler optimizations perform a massive amount of intra- and inter-procedural code restructuring on a whole-program basis. This raises several important and as of yet unanswered questions about how these optimizations impact the code reuse gadget sets present in optimized binaries. Thus, we are motivated to answer the following questions:

\begin{enumerate}
    \item To what degree do compiler optimizations introduce new CRA gadgets into optimized code?
    \item To what degree do compiler optimizations negatively impact the security of optimized code with respect to CRA gadget sets?
    \item Which specific optimization behaviors cause negative security impacts?
    \item What are the root causes of these negative security impacts?
    \item How can these negative security impacts be mitigated?
    \item What is the performance cost of implementing such mitigations?
\end{enumerate}

\subsection{Summary of Contributions} To answer our first three motivating questions, we developed a systematic data-driven methodology to study compiler optimization behavior. Employing this methodology, we built 1,187 total variants of 20 different benchmark programs using two production compilers, \verb|GCC| and \verb|Clang|. Each variant was built with a different optimization configuration to enable coarse-grained analysis of predefined optimization levels (i.e., O0, O1, O2, O3) and fine-grained analysis that isolates the impacts of individual compiler optimization behaviors. We then analyzed and compared these variants to identify negative security impacts using the Gadget Set Analyzer (GSA)~\cite{gadgetSetAnalyzer}. GSA is a static binary analysis tool that catalogs, filters, classifies, and scores gadgets in variant binaries. Using this information, GSA calculates quantitative and qualitative security-oriented metrics for each gadget set that measure its composition, expressive power, special purpose capabilities, and suitability for exploit creation. GSA then compares these metrics across variants to identify how security is impacted by compiler optimizations.

The results of our study are concerning. Our coarse-grained analysis (Section \ref{section:coarse_grained}) revealed that code reuse gadget sets present in optimized program variants are significantly more useful to an attacker than the set in the program's unoptimized variant. These findings are ubiquitous; we observed significant negative security impacts across all compilers, optimization levels, and benchmarks. Diving deeper, our fine-grained analysis (Section \ref{section:fine_grained}) revealed that these impacts are not localized to a few problematic optimizations. While we did observe several optimizations with specific undesirable behaviors, we also observed that large majority of individual optimizations caused small but measurable negative security impacts with respect to gadget sets. 

To answer our last three motivating questions, we conducted differential binary analysis of our variants to identify the root causes of these negative security impacts. Our analysis revealed three distinct root causes: duplication of gadget producing instructions (GPIs), transformation-induced gadget introduction, and special purpose gadgets introduced explicitly by optimizations. After further study of these phenomena, we formulated potential mitigation strategies for each. 

We implemented our proposed mitigation strategies for GPI duplication and transformation-induced gadget introduction as a suite of five binary recompiler passes that address undesirable optimization behaviors without sacrificing their desirable performance benefits. When used to recompile optimized variants of our benchmark programs, these passes reduced the total number of special purpose gadget types available in the recompiled binaries by 34\% and reduced the number of useful gadgets by an average of 31.8\%. Our passes also reduced the total expressive power of the remaining gadgets in 78\% of cases and successfully reduced 81\% of fully expressive gadget sets in optimized variants to a less than fully expressive level. Performance analysis of our passes shows that these benefits can be obtained with negligible impact to execution speed and code size. Recompiling optimized variants with our passes resulted in an average \textbf{speedup} of 0.2\% and incremental static code size increase of 0.5\% (6.1 kilobytes).

Finally, we evaluated the performance impacts of a potential mitigation strategy for special purpose gadgets introduced explicitly by optimizations. Specifically, we evaluated the common practice of disabling optimizations that produce negative security impacts for \verb|Clang|'s tail-call elimination (TCE) optimization. We observed that disabling TCE increases execution time significantly, by 14\% on average. Our evaluation suggests that this basic strategy is likely cost-ineffective for preventing the introduction of certain special purpose gadget types during optimization. 

Our contributions are organized as follows. In Section~\ref{section:setup} we detail our data-driven study methodology. In Section~\ref{section:study}, we present the results of our study of the impacts of compiler optimizations on code reuse gadget sets. In Section~\ref{section:root_causes}, we identify the root causes of and propose mitigation strategies for the negative security impacts we observed. In Section~\ref{section:mitigation}, we detail our implementation, evaluation, and performance analysis of our proposed mitigation strategies.

\section{Background}

\subsection{Gadget-Based Code Reuse Attacks (CRAs)} CRA methods circumvent code injection defenses (e.g., W$\oplus$X) by chaining existing executable snippets of a vulnerable program called gadgets into an exploit payload. \citet{shacham} proposed the first gadget-based CRA method, Return-Oriented Programming (ROP), to overcome the expressivity limitations of early CRA methods (e.g., return-to-libc) that rely on the malicious execution of an existing library function. ROP has been shown to be Turing-complete if a sufficiently expressive set of gadgets is present in the vulnerable program, meaning an attacker can construct and execute any arbitrary program. Several alternative methods to ROP have been introduced, such as jump- and call-oriented programming (JOP, COP)~\cite{bletsch2011jump, checkoway, sadeghi}.

\subsection{Gadget Types} 
\begin{figure}[ht]
\centering
\begin{minipage}{.4\textwidth}
  \footnotesize
  \centering
  \captionof{table}{GPIs and Exploit Patterns (x86-64)}
  {
\begin{tabular}{|c|l|}
    \hline
    \textbf{Exploit Pattern} & \textbf{GPI Group}  \\
    \hline
    & \verb|retn|; \hspace{0.1cm} \verb|retf|; \\ 
    ROP & \verb|retn <imm>|; \\ 
    & \verb|retf <imm>|;  \\
    \hline
    & \verb|jmp <reg>|; \\ 
    JOP & \verb|jmp [reg]|; \\ 
    & \verb|jmp [reg+off]|; \\
    \hline
    & \verb|call <reg>|; \\ 
    JOP / COP & \verb|call [reg]|; \\ 
    &  \verb|call [reg+off]|;  \\
    \hline
    & \verb|int 0x80|; \\ 
    All (Syscall) & \verb|call PTR 0x10|; \\ 
     & \verb|syscall|; \hspace{0.1cm}  \verb|sysenter|; \\
    \hline
\end{tabular}
}\newline
  \label{tab:GPIs}
\end{minipage}%
\begin{minipage}{.6\textwidth}
  \centering
  \includegraphics[width=0.8\linewidth]{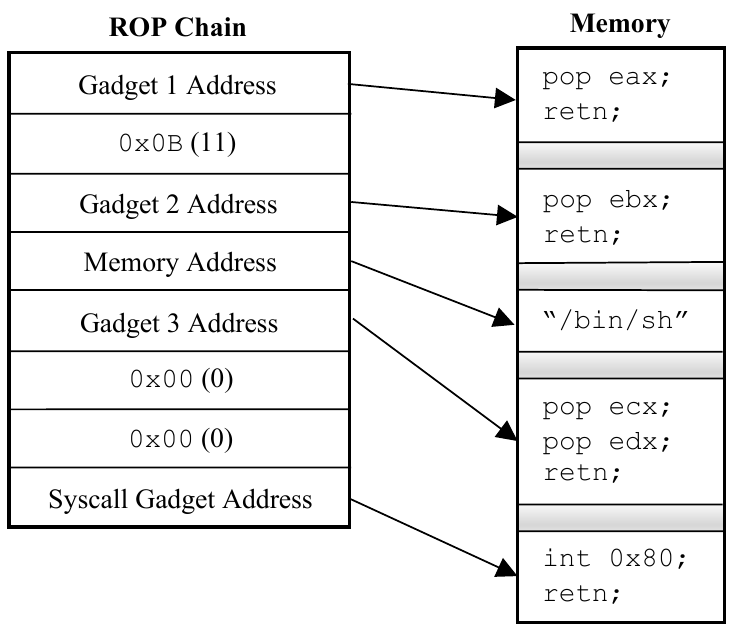}
  \captionof{figure}{Simple ROP Exploit Chain}
  \label{fig:simple-rop}
\end{minipage}
\end{figure}

A CRA gadget is a sequence of machine instructions that ends with one of the control-flow transfer instructions listed in Table \ref{tab:GPIs}, which we refer to as \textbf{Gadget-Producing Instructions (GPIs)}. Not all control-flow transfer instructions are GPIs; specifically direct jump and call instructions are not GPIs. However, conditional and unconditional jumps can be included as one of a gadget's intermediate instructions. Such gadgets are called \textbf{multi-branch gadgets}.

When chained together using the GPI's control flow properties, the gadget sequence is equivalent to an executable program built entirely from existing code segments. Individual gadgets in an exploit chain are used for one of two purposes. \textbf{Functional gadgets} perform computational tasks such as adding values or loading a value into a register. The attacker uses functional gadgets to express their malicious exploit. \textbf{Special purpose gadgets} perform critical non-expressive actions such as invoking system calls (i.e., syscall gadgets) or maintaining control flow in JOP/COP exploits. 

An attacker is not limited to the instructions explicitly generated by the compiler, called \textbf{intended gadgets}, when building exploits. Because \verb|x86-64| instructions are variable length, an attacker can redirect execution to any program byte offset and interpret the byte sequence starting there as a gadget. Due to the density of the \verb|x86-64| encoding space, programs contain large numbers of legal instruction sequences not explicitly generated by the compiler, called \textbf{unintended gadgets}. For example, consider the byte sequence \verb|{83 C0 5B; C3;}|, which encodes the intended gadget \verb|{add eax, 91; ret;}|. If the attacker increments the offset of this gadget by 2 bytes, they can use the unintended gadget \verb|{pop rbx; ret;}| encoded by the sequence \verb|{5B; C3;}|.

\subsection{Exploitation} Gadget-based CRA patterns are typically used to exploit programs with stack-based memory corruption vulnerabilities (e.g., CWE-121). First, the attacker must identify a suitable vulnerability in the target program to exploit, either by confirming the target program contains a publicly-disclosed vulnerability or by discovering a novel vulnerability via methods such as fuzzing, symbolic execution, or binary reverse engineering. Next, the attacker searches the vulnerable program for gadgets and selects those useful for expressing their malicious intent. They then use the selected gadgets to generate a payload that will overflow the vulnerable buffer, write their gadget chain and necessary data to the stack, and overwrite the return address on the stack with the address of the first gadget in the chain. Several automated tools exist that can aid attackers by simplifying or automating this process~\cite{ROPgadget, follnerPshape, 8709204, schwartz}. When delivered, the payload will hijack control-flow and execute the gadgets in sequence. 

Figure \ref{fig:simple-rop} depicts a simple ROP chain that opens a shell. The first element of the ROP chain is the address of a functional gadget in the vulnerable program's memory. It pops an attacker controlled value, 11 (i.e., the identifier for the \verb|execve| system call), from the stack into the \verb|eax| register and returns. The return instruction then transfers control flow to the next gadget via the next address on the stack. This functional gadget loads the address of a string (i.e., \verb|"/bin/sh"|) in program memory into the \verb|ebx| register and returns. In a similar manner, the third gadget loads null values into the \verb|ecx| and \verb|edx| registers. Control-flow then transfers to the final syscall special purpose gadget, which invokes \verb|execve /bin/sh|, opening a shell at the privilege level of the vulnerable program.

\subsection{Defenses}
\label{section:defenses}
Several defensive techniques against gadget-based CRA methods have been proposed and can be generally categorized as compiler-based defenses or binary retrofitting transformations. Both approaches incur increased code size and execution run-time as a trade-off. 

Compiler-based defenses eliminate unintended gadgets by rewriting instructions that contain GPI encodings into equivalent instructions that do not. They then secure intended gadgets by rewriting GPIs to use alternate control-flow mechanisms~\cite{li} or by inserting run-time protections such as encrypted branch targets or control-flow locks to protect them~\cite{onarlioglu, bletsch2011mitigating}. These approaches incur significant performance degradation with average slowdowns in the 5\%-15\% range and code size increases of $\approx$26\%. These impacts are primarily a factor of the number of GPIs in the program; programs with fewer GPIs generally incur less overhead to secure.

Binary retrofitting transformations generally operate by protecting control-flow branches/targets with inserted controls that enforce run-time control-flow integrity (CFI) ~\cite{abadi, 10.5555/2337159.2337171, zhang, hawkins} or detect anomalous run-time behavior~\cite{davi2009dynamic, chen, davi2011ropdefender, yao}. Due to the high frequency of branches in code~\cite{mcfarling1986reducing}, fully-precise detection and mitigation of CRAs with these approaches is very expensive. As such, CFI implementations make precision/performance trade-offs that balance their effectiveness with their performance costs. Coarse-grained CFI implementations~\cite{10.5555/2337159.2337171, zhang, hawkins} have performance overheads as low as 3\% on average, however they have been shown to be ineffective due to their lack of precision~\cite{schwartz,davi2014stitching}. More precise (and expensive) fine-grained CFI implementations incur a 16\% slowdown and an 8\% increase in code size on average, yet still have been shown to have serious weaknesses~\cite{conti,evans}. Even worse, gadget-based control-flow bending attacks have been shown to be viable against a hypothetical fully-precise CFI implementation~\cite{carlini2015control}.

\subsection{Transformation-Induced Gadget Introduction} Our prior work has shown that software transformations can significantly change the quantity and quality of gadgets found in a binary \cite{brown}. Changes to source code or intermediate representation (IR) are reflected in the downstream compiler-produced binary, which can significantly alter the composition of intended gadgets. These changes, as well as changes in binary layout that occur as a result of transformation, also have significant impacts on the composition of unintended gadgets. This work has shown that high gadget introduction rates occur even in conservative debloating transformations; on average 39.5\% of the gadgets present in debloated binaries were not present in the original binaries.
\section{Data-Driven Study Methodology}
\label{section:setup}

Characterizing the impact of compiler optimizations on gadget availability and utility is a complex problem. Modern compilers employ a vast number of optimizations, many of which are synergistic and require careful phase ordering to be effective. Individual optimizations are also quite diverse in their features; they have varying goals (e.g., reducing code size vs. reducing execution time), make different trade-offs (e.g., increase code size to reduce execution time), and have different scopes (e.g., intra- vs. inter-procedural). As a result, compilers bundle optimizations into levels (i.e., O0 - O3) tuned for different objectives such as fast compilation or fast execution. In this section, we present the systematic, data-driven study methodology we developed to characterize the collective and individual security-oriented impacts of compiler optimizations. Our approach is similar in nature to coarse- and fine-grained studies introduced in prior work \cite{10.5555/1144431.1144433, 10.1145/301631.301683}.

\subsection{Benchmark Selection and Variant Generation} We selected a total of 20 C/C++ benchmark programs for this study that are diverse in size, complexity, and functionality (see Table~\ref{tab:benchmarks}). We sourced 13 of our benchmarks from the 29 programs in the SPEC 2006 benchmark set. Ten programs were excluded from this set because they were implemented in Fortran, and six were excluded because \verb|Clang| fails when trying to build them. We then selected seven additional common Linux programs and libraries for our study to further increase the size and diversity of our benchmark set.

\begin{table*}
    \footnotesize
    \caption{Study Benchmarks}
    \label{tab:benchmarks}
    {\rowcolors{2}{lightgray!50}{white}
\begin{tabular}{|l l|l l l l|}
    \hline
    \multicolumn{2}{|c|}{\textbf{Common Linux Programs}} & \multicolumn{4}{c|}{\textbf{SPEC 2006 Benchmark Set}} \\
    \hline
    \verb|Bftpd| v5.1  & \verb|git| v2.21.0  & \verb|401.bzip2| & \verb|403.gcc| & \verb|429.mcf| & \verb|433.milc|  \\
    \verb|gzip| v1.10 &  \verb|httpd| v2.4.39  &  \verb|444.namd|  &  \verb|445.gobmk| & \verb|453.povray|  & \verb|456.hmmer| \\
    \verb|libcUrl| v7.65.0 & \verb|liblmdb| v0.9.23  &  \verb|458.sjeng|  & \verb|462.libquantum| & \verb|470.lbm| & \verb|471.omnetpp| \\
    \verb|libsqlite| v3.28.0 &    & \verb|482.sphinx3| &   &  &  \\
    \hline
\end{tabular}
}
\end{table*}

We then used two production compilers, \verb|GCC| v7.4.0 and \verb|Clang| v8.0.1, to build 1,187 different variants of our benchmarks. To conduct a coarse-grained analysis of optimization behavior, we built four variants per benchmark per compiler at optimization levels O0 through O3. While optimization level definitions vary between compilers, they are consistent for \verb|GCC| and \verb|Clang|. Code produced at level O0 is almost entirely \textbf{unoptimized}. Specifying level O1 enables entry-level optimizations that reduce code size and execution time with minimal incremental compile time. At level O2, optimizations that improve execution speed without increasing code size are enabled. This increases compile time, but further reduces the binary's size and execution time over level O1. Specifying level O3 enables almost all optimizations and generally produces the fastest binaries, but they are generally larger and require more compile time versus level O2.

To conduct our fine-grained analysis, we built variants that isolate the behavior of individual optimizations. We configured the compiler to perform the desired optimization and all optimizations at levels below the level for which the compiler includes the desired optimization by default. For example, \verb|GCC|'s store merging optimization is included at level O2. To generate a variant that isolates this behavior, we configured \verb|GCC| to perform all level O1 optimizations and the store merging optimization. This is necessary to obtain realistic variants because specifying a particular optimization level also includes the optimizations performed at lower levels. Additionally, many optimizations must be run after other optimizations have already made a pass on the code to be effective. We have made the complete set of our study binaries publicly available to support further research in this important area (see Section \ref{app:artifacts}). This repository includes a full list of the individual optimizations we generated for this study~\cite{single_opt}.

\subsection{Variant Analysis} To determine how each optimization level or individual optimization impacts CRA gadget sets, we analyzed variants against an appropriate baseline using GSA~\cite{brown}. For coarse-grained analysis, we compared benchmark variants produced at optimization levels O1, O2, and O3 to the benchmark's unoptimized variant (i.e., O0) as the baseline. For fine-grained analysis, we compared each variant isolating an individual optimization to its subordinate optimization level variant. Continuing with our previous example, we compared benchmark variants that isolate \verb|GCC|'s store merging optimization to their respective O1 variants. 

To isolate and clearly identify the effects of optimizations on our benchmarks, we do not include the benchmark's dynamically linked libraries in our analysis. In practice, gadgets in library code can be used by the attacker as they are mapped into the program's memory space at run-time. However, the library-based gadgets available depend on a build processes separate from our benchmark's; thus we consider them to be confounding variables in this study. Further, utilizing library-based gadgets is difficult in practice due to defensive techniques such as ASLR~\cite{aslr}, BlankIt~\cite{porter2020}, and piece-wise compilation/loading~\cite{quach2018debloating} that have been shown to be highly effective at neutralizing library-based gadgets.


For the baseline and each variant, GSA uses ROPgadget~\cite{ROPgadget} to scan the binary and generate a complete catalog of its gadgets, including multi-branch gadgets. Next, GSA filters out duplicates and gadgets that are unusable (e.g., contain privileged instructions, do not generate attacker-controllable values, etc.). We refer to the gadgets that remain after filtering as \textbf{useful gadgets}. The useful gadgets are then analyzed to determine what functionality they provide, if they can be used for a special purpose, and their suitability for use in a gadget chain. This information is used to calculate four security-oriented metrics via baseline-variant gadget set comparison that measure changes caused by transformations: one of which is quantitative and three of which are qualitative~\cite{brown, homescu, follner2016analyzing}. 

\subsection{Gadget Set Metrics} \textbf{Gadget Introduction Rate} is a quantitative metric that measures how the composition of a gadget set changes following optimization. It is calculated as the percentage of gadgets in a variant binary that are not present in the baseline binary. Gadget addresses are not considered for this metric, meaning gadgets that are not semantically altered but are relocated to a new address after optimization are not considered introduced. It is important to note that positive introduction rates do not necessarily indicate an increase in the number of unique gadgets in a set, since optimizations frequently introduce new gadgets by transforming an existing gadget. 

\textbf{Functional Gadget Set Expressivity} is a qualitative measure of the computational tasks that can be performed with a particular set of gadgets (i.e., the set's expressive power). GSA measures functional gadget set expressivity with respect to three \textbf{levels} (listed in increasing order of overall expressive power): practical ROP exploits, ASLR-proof practical ROP exploits, and Turing-completeness. To achieve a particular level of expressivity (i.e., to be fully expressive at that level), a gadget set must contain at least one gadget that satisfies each of the level's required computational classes. The above levels require a total of 11, 35, and 17 computational classes, respectively. For example, to achieve Turing-completeness a gadget set must contain at least one gadget that adds attacker-controlled values, one that can store an attacker-controlled value to memory, one that can be used to implement a conditional branch, etc.

GSA calculates functional gadget set expressivity by scanning each ROP gadget's first instruction to determine if it implements a required computational class for any level. After the scan, the ROP gadget set's final expressivity measure is the total number of classes satisfied for each level. These expressivity measures for the baseline and variant binaries are compared on a per-level basis to determine changes in expressive power. If an optimization increases the number of satisfied computational classes with respect to one or more levels, this is considered an undesirable result.

\textbf{Functional Gadget Set Quality} is a qualitative measure of the overall utility of functional gadgets for constructing exploit chains. GSA assigns each gadget an initial score of 0, and then scans its intermediate instructions for side constraints (e.g., stack pointer manipulations, memory operations, conditional branches) that they impose on exploit construction. For each side constraint, the gadget's score is increased by a value proportional to the difficulty of accounting for the side constraint with another gadget in a chain~\cite{gsa_criteria}. Thus, the difficulty of using a gadget in a chain increases with its quality score. Finally, GSA computes the average quality score for the baseline and variant gadget sets and compares them. If an optimization decreases the resulting gadget set's average quality score, this is considered an undesirable result.

As an example, consider the the ROP gadget \verb|{add eax, ebx; ret;}|. It contains no intermediate instructions and thus has no side constraints giving it a score of 0. By contrast, the ROP gadget: \verb|{sub rsi, rcx; xor rax, rax; mov qword ptr [rdx], rsi; ret;}| has two intermediate instructions that constrain exploit construction. The first intermediate instruction (\verb|xor rax, rax;|) will overwrite the value of the \verb|rax| register if used in a chain, limiting the attacker's set of controllable registers. The second intermediate instruction (\verb|mov qword ptr [rdx], rsi;|) performs a write to a memory location held in the \verb|rdx| register, which may or may not represent a valid memory location in a hijacked execution. This gadget's score is increased by 1.0 and 2.0 for these intermediate instructions respectively, for a total score of 3.0.

\textbf{Special Purpose Gadget Availability} is a qualitative measure of the types of exploit patterns that can be employed with a gadget set. It is calculated by scanning the baseline and variant gadget sets to identify which special purpose gadget types are available, defined as the presence of at least one gadget of a particular type. GSA identifies ten different kinds of special purpose gadgets: syscall gadgets, four different types of JOP-specific gadgets~\cite{checkoway}, and five different types of COP-specific gadgets~\cite{sadeghi}. If an optimization introduces types of special purpose gadgets that were not previously available, this is considered an undesirable result. Conversely, it is desirable for all special purpose gadgets of a particular type to be removed.

\section{Study Results}
\label{section:study}

\subsection{Optimization Induced Gadget Introduction}
\label{sec:gadget_intro}

To answer our first motivating question \{\emph{To what degree do compiler optimizations introduce new CRA gadgets into optimized code?}\}, we analyzed the gadget sets in our benchmark variants to determine the rate at which optimizations introduce new gadgets. Our results are shown in Table \ref{tab:total_gadgets} and indicate that the collective effects of optimizations at levels O1 through O3 have a large impact on gadget sets. In all cases, more than two-thirds of useful gadgets in optimized binaries were newly introduced. We observed these effects for both compilers and all optimization levels. Excluding the smallest benchmark programs (i.e., \verb|429.mcf| and \verb|470.lbm|) the minimum observed  introduction rate was 84\%. Individual optimizations also introduce gadgets at a high rate. Across \verb|GCC| single optimization variants, we observed an average introduction rate of 25.2\%. For \verb|Clang| single-optimization variants, we observed a much higher average rate of 82.5\%.

With respect to total gadget counts, \verb|GCC| O1 and O2 variants generally contained a similar number or fewer gadgets than their unoptimized counterparts. However, 60\% of \verb|GCC| O3 variants contained a significantly higher number of gadgets (i.e., >10\% increase). This is not surprising, as optimizations that perform  code size for speedup trade-offs are included at the O3 level in both compilers. Such optimizations create path-optimized copies of code segments, which are likely to introduce new unique gadgets when the copied segments contain GPIs. Our results were different for \verb|Clang|, however. Significant increases in the total number of useful gadgets were observed for several optimization levels in all benchmarks except \verb|httpd|, \verb|456.hmmer|, and \verb|458.sjeng|.

Our results clearly indicate that compiler optimizations have a significant impact on the quantity and composition of the code reuse gadgets present in their resulting binaries. While this quantitative analysis is insufficient to draw conclusions about the impact of optimizations on security~\cite{brown}, the high rates of introduction observed suggest that significant impacts to gadget set quality are likely. This motivates our qualitative analysis in the following sections.

\begin{table*}
\centering
    \footnotesize
    \caption{Total Quality Gadgets and Introduction Rate}
    \label{tab:total_gadgets}
    {\rowcolors{4}{lightgray!50}{white}
\begin{tabular}{|l|c|c|c|c|c|c|c|c|}
    \hline
     & \multicolumn{4}{c|}{\textbf{GCC}} & \multicolumn{4}{c|}{\textbf{Clang}} \\
    \cline{2-9}
   \multirow{-2}{*}{\textbf{Benchmark}} & \textbf{O0} & \textbf{O1} & \textbf{O2} & \textbf{O3} & \textbf{O0} & \textbf{O1} & \textbf{O2} & \textbf{O3}\\
    \hline
    Bftpd & 387 & 385 (90\%) & 474 (93\%) & 455 (92\%) & 335 & 451 (93\%) & 462 (94\%) & 449 (94\%) \\
    libcUrl & 5289 & 5448 (97\%) & 6414 (98\%) & 6967 (98\%) & 4509 & 6058 (96\%) & 6506 (97\%) & 6825 (97\%) \\
    git & 19472 & \scriptsize{12696 (96\%)} & \scriptsize{13537 (97\%)} & \scriptsize{13228} (97\%) & 13198 & \scriptsize{19177 (98\%)} & \scriptsize{12266 (98\%)} & \scriptsize{12362 (98\%)} \\
    gzip & 551 & 495 (90\%) & 537 (89\%) & 582 (91\%) & 350 & 495 (93\%) & 343 (90\%) & 442 (91\%) \\
    httpd & 4640 & 4363 (93\%) & 4879 (93\%) & 5392 (93\%) & 4229 & 4368 (92\%) & 4432 (92\%) & 4478 (92\%) \\
    liblmdb & 448 & 561 (95\%) & 572 (96\%) & 539 (95\%) & 421 & 585 (96\%) & 544 (96\%) & 559 (96\%) \\
    libsqlite & 6393 & 5925 (95\%) & 6665 (96\%) & 6611 (96\%) & 5561 & 7863 (96\%) & 6209 (96\%) & 7007 (97\%) \\
    401.bzip2 & 307 & 301 (91\%) & 319 (92\%) & 410 (94\%) & 294 & 374 (93\%) & 341 (91\%) & 318 (90\%) \\
    403.gcc & 16142 & \scriptsize{13512 (96\%)} & \scriptsize{15139 (97\%)} & \scriptsize{16464 (97\%)} & 9854 & \scriptsize{14588 (97\%)} & \scriptsize{13120 (97\%)} & \scriptsize{13588 (97\%)} \\
    429.mcf & 113 & 151 (83\%) & 134 (80\%) & 136 (80\%) & 113 & 128 (75\%) & 155 (79\%) & 143 (78\%) \\
    433.milc & 682 & 675 (92\%) & 699 (92\%) & 939 (94\%) & 519 & 608 (95\%) & 698 (95\%) & 739 (96\%) \\
    444.namd & 770 & 559 (93\%) & 472 (91\%) & 570 (92\%) & 340 & 737 (95\%) & 875 (97\%) & 954 (96\%) \\
    445.gobmk & 5114 & 5169 (84\%) & 5562 (85\%) & 6006 (86\%) & 4146 & 5596 (85\%) & 5296 (85\%) & 5400 (85\%) \\
    453.povray & 8542 & 6021 (94\%) & 6932 (94\%) & 7278 (95\%) & 4475 & 5928 (95\%) & 5728 (95\%) & 5709 (94\%) \\
    456.hmmer & 2023 & 1942 (96\%) & 2013 (96\%) & 2241 (96\%) & 1568 & 1530 (96\%) & 1568 (96\%) & 1585 (95\%) \\
    458.sjeng & 599 & 697 (94\%) & 722 (95\%) & 847 (94\%) & 508 & 535 (95\%) & 524 (94\%) & 583 (95\%) \\
    462.libquantum & 339 & 290 (90\%) & 320 (91\%) & 349 (92\%) & 220 & 304 (91\%) & 310 (91\%) & 316 (91\%) \\
    470.lbm & 88 & 117 (74\%) & 92 (69\%) & 97 (70\%) & 71 & 111 (78\%) & 108 (77\%) & 111 (78\%) \\
    471.omnetpp & 9379 & 8844 (96\%) & 8620 (97\%) & 8788 (97\%) & 8146 & 10054 (97\%) & 9125 (97\%) & 9162 (97\%) \\
    482.sphinx3 & 1033 & 1058 (96\%) & 1161 (96\%) & 1476 (96\%) & 860 & 1000 (95\%) & 1150 (96\%) & 1246 (96\%) \\
    \hline
\end{tabular}
}

    {\justify The values in each column represent the total number of code reuse gadgets present within the binary. The percentage in the parentheses indicates the gadget introduction rate for that variant, calculated as the percentage of total gadgets present in the optimized variant that are not found in the unoptimized variant.\par}
\end{table*}

\subsection{Coarse-Grained Gadget Set Impacts}
\label{section:coarse_grained}
To answer our second motivating question \{\emph{To what degree do compiler optimizations negatively impact the security of optimized code with respect to CRA gadget sets?}\}, we conducted a coarse-grained analysis of optimization impacts on gadget sets with respect to our three qualitative metrics.

\subsubsection{Functional Gadget Set Expressivity} Table~\ref{tab:expressivity_coarse} contains the functional gadget set expressivity data for our O0 - O3 variants. This data indicates that optimization overwhelmingly results in negative functional gadget set expressivity impacts. We observed an increase in expressive power for at least one expressivity level in 87\% (91 of 105) of optimized variants that were not already fully expressive at level O0. This includes \textbf{all} \verb|Clang| variants and 71\% of \verb|GCC| variants. Further, these increases are frequently significant in magnitude and affect all levels of expressivity. In 37\%  (39 of 105) of variants that were not already fully expressive, newly introduced gadgets satisfied 25\% or more of the total computational classes required for at least one expressivity level. In 39\% (41 of 105) of variants that were not already fully expressive, optimization increased expressivity across all three levels.

\begin{table*}
    \footnotesize
    \caption{Coarse-Grained Functional Gadget Set Expressivity}
    \label{tab:expressivity_coarse}
    {\rowcolors{4}{lightgray!50}{white}
\begin{tabular}{|l|c|c|c|c|c|c|c|c|}
    \hline
    & \multicolumn{4}{c|}{\textbf{GCC}} & \multicolumn{4}{c|}{\textbf{Clang}} \\
     \cline{2-9}
   \multirow{-2}{*}{\textbf{Benchmark}} & \textbf{O0} & \textbf{$\Delta$O1} & \textbf{$\Delta$O2} & \textbf{$\Delta$O3} & \textbf{O0} & \textbf{$\Delta$O1} & \textbf{$\Delta$O2} & \textbf{$\Delta$O3} \\
    \hline
    Bftpd &  8/27/9 & (-1/-6/-2) & (-1/-6/0) & (0/-6/-1) & 7/16/5 & \textbf{(1/10/4)} & (0/\textbf{10/5}) & \textbf{(1/5/3)} \\ 
    libcUrl & 9/34/17 & (0/\textbf{1}/-1) & (0/0/-1) & (0/-1/-1) & 7/27/9 & \textbf{(3/7/6)}  & \textbf{(1/4/4)}  & \textbf{(2/5/7)} \\  
    git & 11/35/17 & (0/0/0) &  (-1/-1/0) &  (0/0/0) & 11/35/16 &  (-1/-1/\textbf{1})  & (-1/0/\textbf{1})  & (0/0/\textbf{1}) \\ 
    gzip & 8/24/9 &  (-1/-2/-1) &  (-1/\textbf{4/3}) & (-2/\textbf{2/3}) & 6/11/5 &  \textbf{(2/18/7)} & \textbf{(1/12/3)} & (0/\textbf{11/3}) \\ 
    httpd & 9/34/17 & (\textbf{1/1}/-1) & (\textbf{1}/0/0) &  (0/-1/0) & 8/31/15 &  (\textbf{1/2}/0) & \textbf{(1/3/1)} &  \textbf{(1/2/2)} \\  
    liblmdb & 6/13/5 &  \textbf{(3/5/3)} &  \textbf{(1/9/3)} & \textbf{(1/11/3)} & 5/10/4 &  \textbf{(3/18/7)} & \textbf{(3/18/7)} & \textbf{(2/16/6)} \\
    libsqlite & 10/35/17 & (\textbf{1}/0/0) & (\textbf{1}/0/-1) & (\textbf{1}/0/-2) & 8/33/16 &  (\textbf{2}/0/0) &  (\textbf{2}/0/\textbf{1}) & (\textbf{3/1}/0) \\ 
    401.bzip2 & 7/12/7 &  (-1/\textbf{10}/0) & \textbf{(1/16/2)} & \textbf{(1/15/3)} & 6/10/5 &  \textbf{(1/18/5)} & \textbf{(1/18/7)} & \textbf{(1/16/5)} \\
    403.gcc & 11/35/17 & (-2/-1/0) &  (-1/-1/0) &  (-1/-1/0) & 9/32/16 & (0/\textbf{2/1}) &  (0/\textbf{2/1}) &  (0/\textbf{2/1}) \\ 
    429.mcf & 7/12/5 & (-1/0/-1) &  (0/\textbf{9/1}) &  (0/\textbf{9/1}) & 5/11/5 &  (0/\textbf{4}/0) & (0/\textbf{4}/0) & (0/\textbf{5/1}) \\  
    433.milc & 9/30/12 & (-1/\textbf{1/1}) &  (0/0/\textbf{3}) & (-1/-4/-1) & 6/22/5 & \textbf{(1/5/3)} & \textbf{(1/5/4)} &  \textbf{(2/5/5)} \\  
    444.namd & 7/25/8 &  (0/-2/0) & (\textbf{1}/-4/0) &  \textbf{(1/3/2)} & 7/24/9 &  (0/\textbf{6/6}) &  \textbf{(1/3/4)} & (0/\textbf{2/3}) \\  
    445.gobmk & 9/34/17 & (\textbf{1}/0/0) &  (0/-1/0) &  (\textbf{2/1}/0) & 10/34/17 &  (\textbf{1/1}/0) &  (0/\textbf{1}/0) &  (0/\textbf{1}/0) \\  
    453.povray & 11/35/17 &  (0/-1/-1) &  (-2/-2/-1) &  (-2/-1/-1) & 8/32/15 &  \textbf{(1/1/2)} &  \textbf{(1/2/1)} &  \textbf{(1/2/1)} \\  
    456.hmmer & 9/34/15 &  (0/0/-1) &  (0/-4/-2) &  (0/-2/0) & 6/28/7 &  \textbf{(2/5/8)} &  \textbf{(2/5/9)} &  \textbf{(3/6/8)} \\  
    458.sjeng & 9/24/10 &  (-2/\textbf{2/3}) &  (-1/\textbf{2/1}) &  (-1/\textbf{7/5}) & 7/15/7 &  \textbf{(2/14/4)} &  \textbf{(1/14/5)} &  \textbf{(1/16/6)} \\
    462.libquantum & 8/24/7 &  (-1/\textbf{2}/-1) &  (-1/\textbf{2}/-1) &  (0/\textbf{3/1}) & 6/20/5 & (0/\textbf{3/4}) &  \textbf{(1/7/6)} & \textbf{(1/5/5)} \\  
    470.lbm & 7/18/5 &  (-1/\textbf{1/1}) &  (-1/\textbf{3/3}) &  (-1/\textbf{2/1}) & 6/9/4 & (-1/\textbf{11/1}) &  (0/\textbf{10/1}) &  (-1/\textbf{7}/0) \\  
    471.omnetpp & 11/35/17 &  (0/0/-1) &  (-1/0/-1) &  (-1/0/0) & 11/35/17 &  (0/0/0) &  (0/0/0) &  (0/-1/0) \\ 
    482.sphinx3 & 8/25/11 &  \textbf{(1/6/3)} &  (0/\textbf{3}/0) &  (0/\textbf{4/1}) & 9/27/11 &  (-1/\textbf{2}/0) &  (0/\textbf{5/2}) &  (-1/\textbf{3/3}) \\ 
    \hline
\end{tabular}
}

    \vspace{1ex}
    {\justify Expressivity is expressed as a 3-tuple in which each integer indicates the number of satisfied computational classes with respect to practical ROP exploits, ASLR-proof practical ROP exploits, and Turing-completeness, in that order. For example, a value of 5/5/5 in an O0 column indicates that the unoptimized binary's gadget set contains gadgets that perform five computational tasks required for each expressivity level. Columns marked with the $\Delta$ symbol contain data in parentheses that indicates the difference between the unoptimized and optimized variants (i.e., O[1,2,3] - O0). Positive values (depicted in \textbf{bold} text) in these columns indicate an undesirable outcome: increased expressive power over the unoptimized binary. For example, a value of \textbf{(1/1/1)} in a $\Delta$ column indicates that optimization increased the number of computational tasks performed by the binary's gadget set by one with respect to each expressivity level.\par}
\end{table*}

\subsubsection{Functional Gadget Set Quality} Table~\ref{tab:quality_coarse} contains the functional gadget set quality data for our O0 - O3 variants. This data indicates compiler optimizations have impacts on functional gadget set quality data that are similar in frequency and severity to functional gadget set expressivity. In 97\% (116 of 120) of variants, optimization increased the number of useful gadgets and/or decreased the gadget set's average quality score. We observed significant negative impacts in 74\% (89 of 120) of variants, defined as a >10\% increase in the number of useful gadgets and/or a decrease in average quality of 0.3 or greater. It is worth noting that significant negative impacts occur for all but four \verb|Clang| variants and occur with less frequency in O1 and O2 \verb|GCC| variants.

\begin{table*}
    \footnotesize
    \caption{Coarse-Grained Functional Gadget Set Quality}
    \label{tab:quality_coarse}
    {\rowcolors{4}{lightgray!50}{white}
\begin{tabular}{|l|c|c|c|c|c|c|c|c|}
    \hline
    & \multicolumn{4}{c|}{\textbf{GCC}} & \multicolumn{4}{c|}{\textbf{Clang}} \\
    \cline{2-9}
   \multirow{-2}{*}{\textbf{Benchmark}} & \textbf{O0} & $\Delta$\textbf{O1} & $\Delta$\textbf{O2} & $\Delta$\textbf{O3} & \textbf{O0} & $\Delta$\textbf{O1} & $\Delta$\textbf{O2} & $\Delta$\textbf{O3} \\
    \hline
    
    Bftpd &  382/1.6 & (\textbf{1}/0.3) & (\textbf{80}/0.2) & (\textbf{58}/0.1) &  331/2.0 &  \textbf{(108/-0.4)}  & \textbf{(118/-0.4)} & \textbf{(104/-0.3)} \\  
    libcUrl & 5218/1.7 & (\textbf{155}/0.1) &  (\textbf{1096}/0) & (\textbf{1648}/0)  & 4448/1.7 & \scriptsize{(\textbf{1519}/0.1)} &  (\textbf{1954}/0) &  (\textbf{2302}/0) \\ 
    git & \scriptsize{19137/1.9} & \scriptsize{(-6645/\textbf{-0.1})} &  \scriptsize{(-5823/\textbf{-0.1})} & \scriptsize{(-6120/\textbf{-0.2})} &  \scriptsize{12849/2.1} &  \scriptsize{\textbf{(5945/-0.3)}} &  (-770/\textbf{-0.3}) & (-674/\textbf{-0.3}) \\  
    gzip & 545/1.9 &  (-55/\textbf{-0.2}) &  (-11/\textbf{-0.3})  & \textbf{(30/-0.3)} & 345/1.8 & (147/\textbf{-0.1}) &  (-7/\textbf{-0.4}) &  \textbf{(90/-0.3)} \\  
    httpd & 4534/1.8 &  (-241/\textbf{-0.1}) & \textbf{(272/-0.1)} & \textbf{(767/-0.1)} & 4141/1.9 &  \textbf{(136/-0.2)} &  \textbf{(206/-0.2)} &  \textbf{(245/-0.2)} \\ 
    liblmdb & 446/1.9 &  \textbf{(110/-0.3)} & \textbf{(125/-0.3)} &  \textbf{(91/-0.3)} & 419/1.8 & \textbf{(163/-0.2)} &  \textbf{(124/-0.1)} &  \textbf{(136/-0.1)} \\  
    libsqlite & 6294/1.8 & (-483/\textbf{-0.1}) & (\textbf{243}/0) & \textbf{(218/-0.1)} & 5472/1.7 &  (\textbf{2245}/0) &  (\textbf{635}/0) &  \scriptsize{\textbf{(1421/-0.1)}} \\  
    401.bzip2 & 306/1.7 &  (-6/\textbf{-0.1}) &  (\textbf{10}/0) &  \textbf{(100/-0.1)} & 290/2.0 & \textbf{(83/-0.4)} &  \textbf{(49/-0.5)} &  \textbf{(27/-0.5)} \\  
    403.gcc & \scriptsize{15899/2.0} &  \scriptsize{(-2524/\textbf{-0.3})} &  (-989/\textbf{-0.2}) & \textbf{(337/-0.3)} & 9768/2.0 & \scriptsize{\textbf{(4638/-0.3)}} & \scriptsize{\textbf{(3184/-0.3)}} & \scriptsize{\textbf{(3667/-0.3)}} \\
    429.mcf & 112/1.7 & \textbf{(35/-0.3)} &  \textbf{(20/-0.3)} & \textbf{(23/-0.1)} & 112/1.5 &  (\textbf{15}/0) & (\textbf{41}/0.2) &  (\textbf{29}/0.1) \\  
    433.milc & 662/1.6 & (\textbf{7}/0.3) &  (\textbf{32}/0.1) &  (\textbf{264}/0.1) & 511/1.9 & \textbf{(96/-0.4)} & \textbf{(181/-0.4)} &  \textbf{(226/-0.4)} \\ 
    444.namd & 752/2.1 &  (-205/\textbf{-0.2}) & (-287/\textbf{-0.5}) &  (-191/\textbf{-0.5}) & 339/1.6 &  (\textbf{392}/0.2) &  \textbf{(523/-0.1)} &  (\textbf{596}/0) \\  
    445.gobmk & 4938/1.8 & (\textbf{72}/0.1) & (\textbf{459}/0.1) & (\textbf{858}/0.1) & 3976/2.0 &  \scriptsize{\textbf{(1412/-0.1)}} & \scriptsize{\textbf{(1119/-0.2)}} &  \scriptsize{\textbf{(1227/-0.1)}} \\  
    453.povray & 8273/1.6 &  \scriptsize{(-2348/\textbf{0.2})} &  \scriptsize{(-1522/\textbf{0.1})} &  \scriptsize{(-1149/\textbf{0.1})} & 4407/1.7 & \scriptsize{\textbf{(1414/-0.1)}} &  \scriptsize{\textbf{(1220/-0.2)}} &  \scriptsize{\textbf{(1198/-0.1)}} \\  
    456.hmmer & 1972/2.0 & (-59/\textbf{-0.3}) &  (0/\textbf{-0.3}) &  \textbf{(237/-0.3)}  & 1544/2.1 & (-31/\textbf{-0.5}) & \textbf{(12/-0.5)} &  \textbf{(15/-0.5)} \\  
    458.sjeng & 588/1.5 &  (\textbf{104}/0.4) &  (\textbf{128}/0.4) & (\textbf{251}/0.3) & 501/2.0 &  \textbf{(31/-0.4)} & \textbf{(19/-0.1)} &  \textbf{(79/-0.1)} \\  
    \scriptsize{462.libquantum} & 331/1.9 &  (-42/\textbf{-0.5}) &  (-13/\textbf{-0.5}) & \textbf{(15/-0.5)} & 217/1.8 &  \textbf{(85/-0.2)} &  \textbf{(90/-0.2)} &  \textbf{(98/-0.1)} \\  
    470.lbm & 86/1.4 &  \textbf{(30/-0.1)} & (\textbf{4}/0) & \textbf{(10/-0.1)} & 69/1.7 &  \textbf{(39/-0.3)}  & \textbf{(38/-0.2)}  & \textbf{(40/-0.2)} \\ 
    471.omnetpp & 9151/2.0 & (-511/0) &  (-875/\textbf{-0.1}) & (-682/\textbf{-0.1}) & 8000/2.1 & \scriptsize{\textbf{(1734/-0.2)}} & \textbf{ (868/-0.1)} & \textbf{(930/-0.1)} \\ 
    482.sphinx3 & 1018/1.8 & (\textbf{34}/0) & \textbf{(133/-0.1)} & \textbf{(448/-0.1)} & 852/1.7 & (\textbf{135}/0.1) &  (\textbf{287}/0) &  \textbf{(375/-0.1)} \\ 
    \hline
\end{tabular}
}

    \vspace{1ex}
    {\justify Gadget set quality is expressed as a 2-tuple in which the first integer indicates the number of useful functional gadgets in the variant and the second value indicates their average quality score. For example, a value of 1000/1.0 in an O0 column indicates that the unoptimized binary's gadget set contains 1000 useful gadgets and the average quality score for the set is 1.0. Columns marked with the $\Delta$ symbol contain data in parentheses that indicates the difference between the unoptimized and optimized variants (i.e., O[1,2,3] - O0). Positive values (depicted in \textbf{bold} text) for the count indicates an undesirable effect: an increase in the number of the useful gadgets. The reverse is true for the second value; negative values (depicted in \textbf{bold} text) here indicate a decrease in the average difficulty required to chain the gadgets in the variant. For example, a value of \textbf{(100/-0.5)} in a $\Delta$ column indicates that optimization increased the number of useful gadgets by 100 and decreased the average gadget set quality score by 0.5 (indicating there are fewer average side constraints per gadget).\par}
\end{table*}

\subsubsection{Special Purpose Gadget Availability} Table ~\ref{tab:special_purpose_gadgets} contains the special purpose gadget availability data for our O0 - O3 variants. We observed overall negative impacts on special purpose gadget availability in 22\% (26 of 120) of total variants and overall positive impacts in 28\% (34 of 120) of total variants. We did not observe overall changes in the other half of total variants, due in part to the relatively small population of special purpose gadgets in our benchmarks and the calculation method for this metric. An overall result of no change is possible in situations where the number of gadget types eliminated is offset by introduction of an equal number of other types. A detailed analysis of the specific types of gadgets present in our variants revealed that optimizations introduced at least one new type of special purpose gadget category that was not originally present in 33\% (39 of 120) of variants. 

\begin{table*}
    \footnotesize
    \caption{Coarse-Grained Special Purpose Gadget Availability}
    \label{tab:special_purpose_gadgets}
    {\rowcolors{4}{lightgray!50}{white}
\begin{tabular}{|l|c|c|c|c|c|c|c|c|}
    \hline
    & \multicolumn{4}{c|}{\textbf{GCC}} & \multicolumn{4}{c|}{\textbf{Clang}} \\
     \cline{2-9}
   \multirow{-2}{*}{\textbf{Benchmark}} & \textbf{O0} & $\Delta$\textbf{O1} &  $\Delta$\textbf{O2} & $\Delta$\textbf{O3} & \textbf{O0} & $\Delta$\textbf{O1} &  $\Delta$ \textbf{O2} & $\Delta$ \textbf{O3}\\
    \hline
    Bftpd & 4 & (-3) & (-1) & (0) & 2 & \textbf{(1)} & \textbf{(2)} & \textbf{(1)} \\ 
    libcUrl & 7 & (0) & (-2) & (0) & 7 & (0) & (0) & (-1) \\ 
    git & 7 & (0) & (0) & (0) & 7 & (0) & (0) & (0) \\ 
    gzip & 3 & (-1) & (0) & \textbf{(2)} & 4 & (-2) & (0) & (0) \\ 
    httpd & 7 & (0) & (0) & (0) & 7 & (0) & (-1) & (0) \\ 
    liblmdb & 2 & (0) & (-1) & (0) & 2 & \textbf{(1)} & (-1) & \textbf{(1)} \\ 
    libsqlite & 7 & (0) & (0) & (0) & 7 & (0) & (-1) & (0) \\ 
    401.bzip2 & 1 & (0) & \textbf{(2)} & \textbf{(1)} & 3 & (-2) & (-1) & (-2) \\ 
    403.gcc & 7 & (0) & (0) & (0) & 7 & (0) & (0) & (0) \\ 
    429.mcf & 1 & \textbf{(1)} & \textbf{(1)} & (0) & 1 & (0) & \textbf{(1)} & (0) \\ 
    433.milc & 5 & (-2) & (-2) & (0) & 4 & (-3) & (-2) & (-2) \\ 
    444.namd & 3 & \textbf{(1)} & (0) & \textbf{(1)} & 1 & \textbf{(3)} & \textbf{(2)} & \textbf{(2)} \\ 
    445.gobmk & 6 & (0) & \textbf{(1)} & \textbf{(1)} & 7 & (-1) & (0) & (0) \\ 
    453.povray & 7 & (0) & (0) & (0) & 7 & (-1) & (-1) & (0) \\ 
    456.hmmer & 6 & (-1) & (-1) & (-1) & 4 & \textbf{(1)} & (0) & \textbf{(2)} \\ 
    458.sjeng & 3 & \textbf{(2)} & (0) & \textbf{(1)} & 3 & (-1) & (0) & (0) \\ 
    462.libquantum  & 3 & (-2) & (-1) & (0) & 3 & (-1) & (-1) & (-2) \\ 
    470.lbm & 2 & (-1) & (0) & (-1) & 1 & \textbf{(1)} & (0) & (0) \\ 
    471.omnetpp & 7 & (-1) & (0) & (0) & 7 & (0) & (0) &(0) \\ 
    482.sphinx3 & 4 & (0) & (0) & \textbf{(1)} & 4 & (-1) & \textbf{(1)} & \textbf{(1)} \\ 
    \hline
\end{tabular}
}
    \vspace{1ex}
    {\justify Special purpose gadget availability is expressed as a single integer indicating the number of special purpose gadget categories available in the gadget set. A category is considered available if at least one special purpose gadget of that type is present. For example, a value of 5 in an O0 column indicates that the unoptimized binary's gadget set has 5 types of special purpose gadget types available. In columns marked with the $\Delta$ symbol, the integer in parentheses indicates the difference between the unoptimized and the optimized variants (i.e., O[1,2,3] - O0). Positive values (depicted in \textbf{bold} text) in these columns indicate an undesirable result: an increase in the number of special purpose gadget types available in the optimized variant's gadget set. For example, a value of \textbf{(2)} in a $\Delta$ column indicates that optimization increased the number of special purpose gadget types available by 2 categories.\par}
\end{table*}

The most frequently introduced special purpose gadgets were syscall gadgets, JOP dispatcher gadgets, and COP dispatcher gadgets. Syscall gadgets are particularly dangerous and versatile special purpose gadgets that can be used in all exploit patterns for performing sensitive actions such as executing programs or opening shells. The introduction of these critical special purpose gadgets enables relatively simple and powerful exploits such as the one depicted in Figure \ref{fig:simple-rop}. Dispatcher gadgets are similarly indispensable. In JOP and COP exploit patterns, the attacker cannot rely on the stack semantics of the \verb|ret| GPI to chain gadgets together; instead they primarily rely on dispatcher gadgets to maintain control flow. While the attacker can use other special purpose gadgets (e.g., JOP and COP trampolines) to chain functional gadgets, these methods are frail and easier to detect by defensive techniques \cite{10.1145/2366231.2337171, yao}.

\subsubsection{Discussion} Our coarse-grained analysis indicates that the gadget sets available in optimized program variants are significantly more useful for constructing CRA exploits than the sets found in unoptimized variants. This conclusion is consistent with the general separation of concerns in optimizing compilers. Compiler front ends responsible for lowering (i.e., translating) source code to intermediate representation (IR) focus on capturing program semantics in IR, whereas the compiler middle end optimizes the IR in place with a focus on improving performance. Many optimizations achieve performance improvements by transforming the relatively simple IR produced by the front end into IR that is more computationally diverse and complex. Coupled with our prior observation that optimizations both introduce new gadgets at a high rate and increase the total number of gadgets (see Section \ref{sec:gadget_intro}), it follows that optimization behaviors that increase code diversity and complexity also increase the availability and utility of code reuse gadgets found within the optimized code.

Interestingly, our data indicates that negative security impacts do not follow a linear progression, do not show biases or co-relations to optimization levels, and are not significantly impacted by the input program. We observed significant negative impacts across all optimization levels, benchmarks programs, and compilers (more so for \verb|Clang|). We conclude from this observation that \textbf{avoiding negative security impacts on CRA gadget sets is not as simple as selecting a particular optimization level}. 

\subsection{Fine-Grained Gadget Set Impacts}
\label{section:fine_grained}

To answer our third motivating question \{\emph{Which specific optimization behaviors cause negative security impacts?}\}, we conducted a fine-grained analysis of gadget set impacts across 1,027 single optimization variants of our benchmarks. Using the coarse-grained variants produced at levels O0, O1, and O2 as baselines, we analyzed our single optimization variants with GSA to isolate their effects. Similarly to our coarse-grained analysis, we observed that our fine-grained variants exhibited a variety of negative security impacts on gadget sets. Increases in gadget set expressivity, gadget set quality, and special purpose gadget availability were commonly observed in variants across all benchmarks, isolated optimizations, and compilers; however, they were observed with more variability than in our coarse-grained variants. 

In the majority of single optimization variants, negative impacts were relatively small in magnitude, manifesting as gadget set expressivity increases of one class, an increase of less than 5\% in the number of useful gadgets, or the introduction of one type of special purpose gadget.\footnote{Syscall, JOP data loader, and COP intra-stack pivot gadgets were the most frequently introduced special purpose gadgets.} We also observed small magnitude impacts across our single optimization variants that were positive, albeit at a lower incidence rate than negative impacts. Our prior work~\cite{brown} suggests that transformation-induced gadget introduction is responsible for the majority of this observed "background noise". We detail our investigation of this potential root cause in Section \ref{section:transformation_induced_introduction}.

\subsubsection{Outlier Detection} 
Within this "background noise", we observed a number of individual optimizations with negative impacts that were significantly larger in magnitude. To separate this "signal" for deeper analysis, we performed outlier detection across our single optimization variant data. We define outliers as variants with metric result changes from the baseline variant greater than 1.5 times the standard deviation from the mean metric result change on a per benchmark, per metric basis. We then combined the total list of identified outliers across all benchmarks and metrics into a histogram to identify which optimizations produced outliers most frequently. We identified that \verb|GCC|'s \textbf{Interprocedural Constant Propagation (IPA CP) Function Cloning} (\verb|-fipa-cp-clone|) optimization frequently produced outlier impacts with respect to functional gadget set expressivity. We also identified four optimizations that frequently produced outlier impacts with respect to special purpose gadget availability: \verb|GCC|'s \textbf{Jump-Following Common Subexpression Elimination (CSE)} (\verb|-fcse-follow-jumps|), \textbf{Peephole} (\verb|-fpeephole2|), \textbf{Omit Frame Pointer} (\verb|-fomit-frame-pointer|), and \verb|Clang|'s \textbf{Tail Call Elimination (TCE)} (\verb|--tail-call-elim|).

\subsubsection{Outlier Analysis}
Table \ref{tab:single_optimization_variants} contains a subset of our fine-grained analysis data for the optimizations that frequently produce outliers. \verb|GCC|'s IPA CP function cloning optimization was observed to cause negative impacts on functional gadget set expressivity in 55\% of variants, with outlier increases to at least one expressivity level occurring in 20\% of total variants. \verb|GCC|'s jump-following CSE, peephole, and omit frame pointer optimizations frequently introduced new types of special purpose gadgets, most commonly syscall and COP intra-stack pivot gadgets. We observed new categories of special purpose gadgets introduced by these optimizations in 35\%, 45\%, and 35\% of total variants, respectively. In contrast, \verb|Clang|'s TCE optimization had a higher number of positive than negative impacts on special purpose gadget availability, eliminating one or more types of special purpose gadgets in 40\% of total variants. New types of special purpose gadgets were introduced in 20\% of variants. Deeper analysis of this unexpected result revealed that this optimization frequently eliminates COP-specific special purpose gadgets at the cost of introducing new JOP data loader gadgets at a high rate. The ultimate effect of the optimization (i.e., positive or negative) largely depends on the types of special purpose gadgets present in the baseline variant. For example, baseline variants without JOP data loader or COP-specific gadget types are likely to suffer the undesirable impact of newly introduced JOP data loader gadgets without a corresponding benefit of COP-specific gadget elimination. 

\begin{table*}
    \footnotesize
    \caption{Single Optimization Variants with Frequent Outliers}
    \label{tab:single_optimization_variants}
    {\rowcolors{4}{lightgray!50}{white}
\begin{threeparttable}
\begin{tabular}{|l|c|c|c|c|c|c|c|c|c|c|}
    \hline
     & \multicolumn{2}{c|}{\textbf{IPA CP Clone*} } 
     & \multicolumn{2}{c|}{\textbf{Jump Follow CSE}} 
     & \multicolumn{2}{c|}{\textbf{TCE}} 
     & \multicolumn{2}{c|}{\textbf{Peephole}} 
     & \multicolumn{2}{c|}{\textbf{Omit Frame Ptr.}}\\
      \cline{2-11}
   \multirow{-2}{*}{\textbf{Benchmark}} & \textbf{O2} & $\Delta$\textbf{Opt} & \textbf{O1} & $\Delta$\textbf{Opt} & \textbf{O0} & $\Delta$\textbf{Opt} & \textbf{O1} & $\Delta$\textbf{Opt} & \textbf{O0} & $\Delta$\textbf{Opt} \\
    \hline
    Bftpd & 7/21/9 & (0/-1/\textbf{3}) & 1 &  (0) & 2 &  \textbf{(1)} & 1 & (0) & 4 & \textbf{(1)} \\
    libcUrl  & 9/34/16 &  (0/0/0) & 7 & (0) & 7 & (-1) & 7 & (-1) & 7 & (0)\\
    git  & 10/34/17  & (0/0/0) & 7 &  (0) & 7 & (0) &  7 & (0)& 7 & (0) \\
    gzip  &  7/28/12  & (-1/-2/\textbf{1}) & 2 & \textbf{(2)} & 4 &  (-1) & 2  & (0) & 3 & (0) \\
    httpd & 10/34/17 & (0/0/-1) & 7 &  (0) & 7 &  (-1) & 7 &  (0) & 7 & (0) \\
    liblmdb  & 7/22/8  & \textbf{(1/2/1)} & 2 & (0) &  2 & \textbf{(1)} & 2 & (0)& 2 & (0) \\
    libsqlite  & 11/35/16  & (-1/-1/0) & 7 & (0) & 7 & (0) & 7 & (0) & 7 & (0) \\
    401.bzip2  & 8/28/9  & (0/\textbf{1}/0) & 1 & (0) & 3 & (-1) & 1 & \textbf{(1)} & 1 & \textbf{(2)}\\
    403.gcc & 10/34/17  & (0/\textbf{1}/0) & 7 & (0) & 7 & (0) & 7 & (0) & 7 & (0) \\
    429.mcf  & 7/21/6 &  (0/0/0) & 2 & (0) & 1 & \textbf{(1)} & 2 & (-1) & 1 & (0) \\
    433.milc  & 9/30/15 &  (-1/-1/-2) & 3 & \textbf{(1)} & 4 & (-1) & 3 & \textbf{(2)} & 5 & (-1) \\
    444.namd  & 8/21/8 & (0/0/0) & 4 & \textbf{(2)} & 1 & (0) & 4  & \textbf{(1)} & 3& \textbf{(1)} \\
    445.gobmk & 9/33/17 &  (\textbf{2/2}/-1) & 6 & \textbf{(1)} & 7 & (0) & 6  & \textbf{(1)} & 6 & \textbf{(1)} \\
    453.povray  & 9/33/16 &  \textbf{(1/1/1)} & 7 & (0) & 7 & (-1) & 7 & (0) & 7 & (0) \\
    456.hmmer & 9/30/13 &  (-1/\textbf{1/1}) & 5 & \textbf{(1)} & 4 & \textbf{(2)} & 5 & \textbf{(1)} & 6 & (0) \\
    458.sjeng  & 8/26/11 &  (0/\textbf{3/2}) & 5 & (-3) & 3 & (0) & 5 & (-3)& 3 & \textbf{(1)}\\
    462.libquantum & 7/26/6 &  (0/0/0) & 1 & \textbf{(2)} & 3 & (-1) & 1 & \textbf{(1)} & 3 & \textbf{(1)} \\
    470.lbm  & 6/21/8 &  (0/0/0) & 1 & (0) & 1 & (0)  & 1 & \textbf{(1)} & 2 & (0) \\
    471.omnetpp & 10/35/16  & (0/-1/\textbf{1}) & 6 & \textbf{(1)} & 7 & (0) & 6 & \textbf{(1)} & 7 & (-1) \\
    482.sphinx3 & 8/28/11 & (0/\textbf{1}/0) & 4 & (0) & 4 & (-1) & 4 & \textbf{(1)} & 4 & \textbf{(1)} \\
    \hline
\end{tabular}
\begin{tablenotes}
      \item * Functional gadget set expressivity data is displayed for this variant. For all others, special purpose gadget availability data is displayed.
    \end{tablenotes}
\end{threeparttable}
}
\end{table*}

\subsubsection{Discussion} We draw two high-level conclusions from our fine-grained analysis. First, nearly all optimizations have a small but measurable impact on the composition of the resulting gadget set. The majority of these impacts are negative, though positive impacts were also observed at lower frequency. This conclusion is consistent with the findings of our coarse-grained analysis; it suggests that the high levels of negative impacts observed at predefined optimization levels are the conglomeration of smaller magnitude impacts made by many individual optimizations rather than a small group of troublesome optimizations with high magnitude impacts. Second, we conclude that a relatively small number of optimizations have impacts that are clearly discernible from the "background noise". This conclusion is promising as it suggests that there are relatively few undesirable optimization behaviors that are negatively impacting gadget sets.
\section{Root Causes and Mitigation Strategies}
\label{section:root_causes}

To answer our fourth motivating question \{\emph{What are the root causes of these negative security impacts?}\}, we performed differential binary analysis of our single optimization variants and their respective baseline variants to determine the underlying causes for the negative impacts we observed. While we focused our efforts on the optimizations identified in Section \ref{section:fine_grained}, we analyzed a representative sample of other optimizations in order to confirm our hypothesis that transformation-induced gadget introduction is the underlying cause of the small magnitude gadget set impacts we observed. We used IDA Pro and BinDiff~\cite{ida, bindiff} to identify and inspect the after-effects of individual optimizations. BinDiff uses heuristics to match functions between different versions of the same program and provides a visual before-and-after comparison of the control flow graphs (CFGs) for each version. To ensure precise CFG recovery during disassembly, we built our program variants with full debug information. Through differential analysis, we identified three root causes of the negative gadget set impacts we observed, which we detail in the following subsections. To address our fifth motivating question \{\emph{How can these negative security impacts be mitigated?}\}, we also propose potential mitigation strategies for each root cause in this section. 

\subsection{GPI Duplication}

Gadget search algorithms scan binaries for byte sequences that encode GPIs, including both byte sequences intentionally inserted by the compiler (e.g., return instructions terminating functions) as well as sequences that are unintentionally encoded in data, memory addresses, constants, displacements in control-flow instructions, etc. For each GPI byte sequence found, the algorithm attempts to disassemble the bytes preceding the GPI to determine if they encode valid instructions. If successful, the algorithm catalogs the valid sequence of bytes and the GPI encoding as a gadget. This is done iteratively, with each iteration attempting to disassemble a longer byte prefix to the GPI until the length of the sequence exceeds some useful threshold (e.g., 10 bytes). This process identifies all intended and unintended gadgets associated with each GPI encoding at or below the threshold, which collectively forms the binary's gadget set. When an optimization duplicates a GPI, a new set of intended and unintended gadgets associated with the GPI is created. Due to the density of the \verb|x86-64| instruction set and its support for variable length instructions, this set is likely to contain new, unique, and useful gadgets. 

A number of compiler optimizations typically employed at the O3 level selectively duplicate code in order to improve performance at the cost of increased code size. Such behavior duplicates GPIs at a high rate, however GPI duplication also occurs at lower optimization levels. For example, both \verb|GCC| and \verb|Clang| perform function inlining at level O2 and above, which replaces function call sites with the body of the called function. This enables intra-procedural optimizations across caller and callee code, but also duplicates GPIs present in the inlined function's body. This behavior partially explains the negative impacts we observed in our coarse-grained analysis at levels O2 and O3.

Interestingly, our analysis revealed that this behavior also occurs with a number of \verb|GCC|-specific optimizations at the O1 level. GPI duplication was most apparent for \verb|GCC|'s omit frame pointer optimization, which identifies functions that do not require a frame pointer and eliminates pointer setup and restore instructions from the function prologue and epilogue, respectively.\footnote{We did not observe this behavior in Clang due to differences in how it generates function epilogues.} Due to \verb|GCC|'s code generation conventions, the pointer restore instruction (i.e., \verb|pop rbp;|) is typically the lone instruction executed prior to a function's \verb|retn| instruction. When multiple code paths converge at the end of a function, eliminating the pointer restore instruction allows the \verb|retn| instruction to be hoisted to the end of each converging path. While this technique slightly reduces code size and execution time per path by replacing a 5-byte \verb|jmp| instruction with a copy of the single byte \verb|retn| instruction it targets, it duplicates the GPI one or more times. 

\subsubsection{Concrete Example} 
\begin{figure}[ht]
  \centering
  \includegraphics[width=\linewidth]{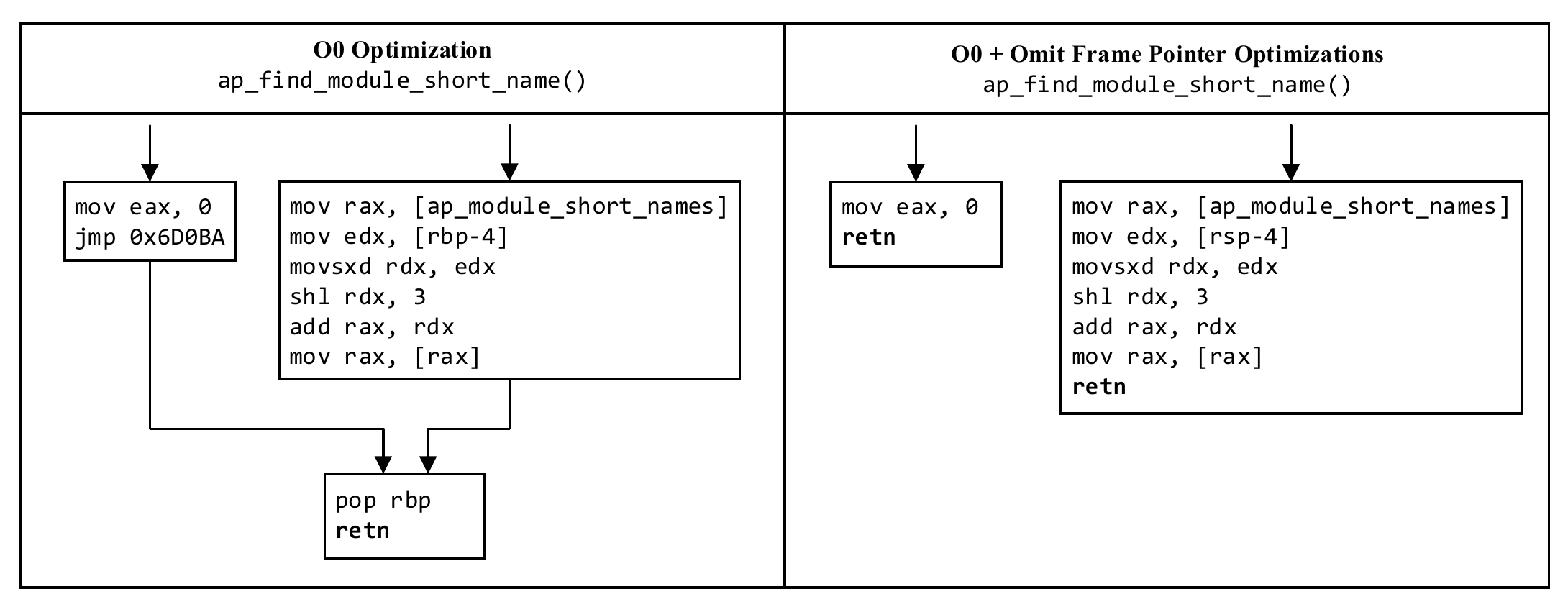}
  \caption{GPI Duplication caused by GCC's Frame Pointer Omission Optimization}
  \label{fig:httpd-omit-frame-ptr}
\end{figure}

An example of this phenomenon in \verb|httpd| is shown in Figure~\ref{fig:httpd-omit-frame-ptr}. Before optimization, the single GPI produces two sets of gadgets: one consisting of intended and unintended gadgets sourced from the instruction sequence \verb|{mov rax, [ap_module_short_names]; ...;| \verb|pop rbp; retn;}| and one consisting of mostly intended gadgets sourced from the multi-branch instruction sequence \verb|{mov eax, 0; jmp 0x6D0BA; pop rbp; retn;}|. The second set contains few, if any, unintended gadgets because the \verb|jmp| in this multi-branch sequence must be executed as encoded to reach the GPI. After the optimization eliminates the \verb|pop rbp;| instruction and duplicates the GPI, the second sequence of instructions is no longer constrained to intended execution as the \verb|jmp| instruction has been eliminated (i.e, it is no longer a multi-branch sequence). The net result is a drastic increase in the total number of unintended gadgets sourced from the two GPIs, as well as significant changes to the composition of the intended gadgets. This behavior is not limited to \verb|GCC|'s omit frame pointer optimization; it also occurs with \verb|GCC|'s shrink wrap and tree switch conversion optimizations. Additionally, this behavior can occur with non-\verb|retn| GPIs. Although rare, we observed instances in which indirect jump and call GPIs were duplicated by \verb|GCC|'s sibling call optimizations. 

\subsubsection{Mitigation Strategy} The negative effects of GPI-duplicating optimizations occur as a result of aggressive secondary optimization (i.e., GPI hoisting). One potential mitigation strategy is to patch compiler optimization passes to avoid this undesirable behavior, eliminating negative gadget set impacts while still obtaining the primary intended benefits of the original optimization. Since several distinct optimizations exhibit this behavior, we choose to evaluate this mitigation strategy in Section~\ref{section:postprod_merging} with binary recompiler passes that intraprocedurally merge GPIs rather than patch multiple compiler passes and \verb|GCC|'s code generation engine.

\subsection{Transformation-Induced Gadget Introduction}
\label{section:transformation_induced_introduction}

Our prior work has shown that even relatively small software transformations introduce a significant number of new gadgets~\cite{brown}. Differential analysis of our single-optimization variants confirmed that seemingly innocuous and localized compiler optimizations are no exception. We refer to this phenomenon as transformation-induced gadget introduction. It is primarily driven by changes  that occur during optimization to program layout and the code preceding compiler-placed GPIs. It results in changes to both intended and unintended gadget populations and is the primary cause of the "background noise" we observed in the large majority of optimized variants in our study.

While there are several mechanisms by which transformation-induced gadget introduction occurs, the most frequent cause we observed was rooted in displacement/offset encodings in control-flow transfer instructions like direct jumps and function calls. Semantic optimizations can induce layout changes to programs, even for simple optimizations such as instruction re-ordering. The resulting changes to displacement/offset encodings can alter a previously benign encoding to one that contains a GPI, which introduces several new unintended gadgets.

\subsubsection{Concrete Example} 
\begin{figure}[ht]
  \centering
  \includegraphics[width=\linewidth]{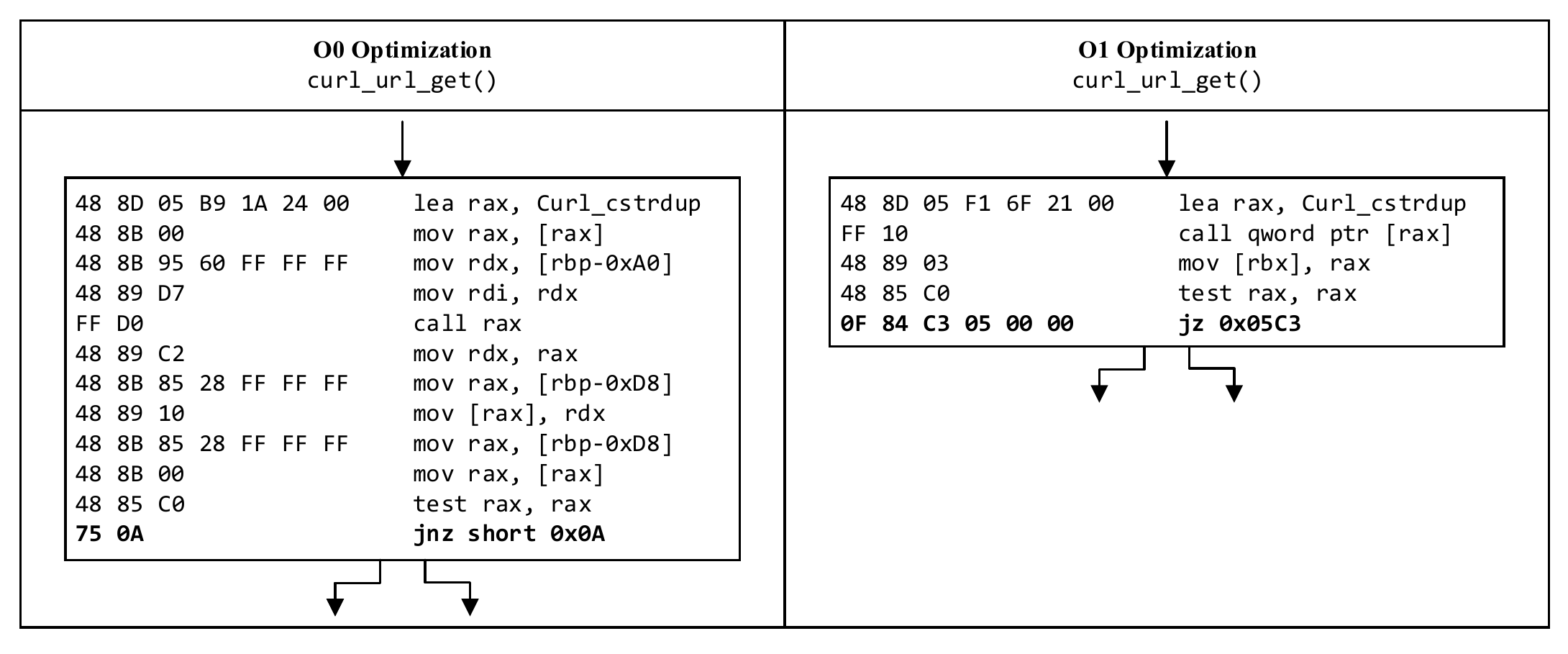}
  \caption{Layout-Based Introduction of GPI and Unintended Gadgets}
  \label{fig:libcurl-rop}
\end{figure}

An example of this behavior in \verb|libcUrl| is shown in Figure~\ref{fig:libcurl-rop}. Here, a conditional jump instruction with a short 1 byte displacement is converted into an equivalent conditional jump instruction with a near 4-byte displacement as a byproduct of optimization. This new displacement encodes the \verb|retn| GPI (i.e., \verb|0xC3|). This behavior is not limited to \verb|retn| GPIs; we also observed two-byte syscall and COP intra-stack pivot special purpose gadgets introduced in this manner, although at a lower rate.

\subsubsection{Mitigation Strategy} Given that this type of gadget introduction is endemic to the \verb|x86-64| instruction set architecture, it is outside the control and concern of machine-independent optimizations. As such, it is not possible to patch them to detect and avoid these harmful effects. However, compiler back-ends responsible for producing the binary can be patched to correct problematic encodings after optimization is complete. Another viable mitigation strategy is post-production transformation passes to eliminate layout-based GPIs. An assembler injection-based solution, G-Free~\cite{onarlioglu}, has been proposed to address this problem, however it is limited to \verb|retn| GPIs and suffers significant performance impacts (average 3.1\% slowdown and 25.9\% increase in code size) due to the lack of direct control over program layout. We propose a new approach to implement this strategy that employs transformations that exercise direct control over program layout. This strategy takes advantage of binary recompilation infrastructure to effectively eliminate unintended gadgets without suffering performance losses. We detail the implementation and evaluation of this mitigation strategy in Section~\ref{section:postprod_merging}.

\subsection{Special Purpose Gadgets Explicitly Introduced by Optimizations}

JOP and COP exploit patterns require special purpose gadgets to perform important non-functional tasks in exploit chains. JOP data loader gadgets pop the attacker's data from the stack into registers at the beginning of a JOP exploit chain. They consist of an indirect jump GPI preceded by a \verb|popa| or multiple \verb|pop| instructions. However, the utility of this instruction sequence is not limited to exploit programming. \verb|Clang|'s tail call elimination (TCE) optimization identifies cases where a function's stack frame can be reused by a child function call occurring at its end. The optimization replaces the tail call and the following return instruction with a jump to the child function, eliminating stack frame set up operations. Indirect call GPIs eliminated in this manner are replaced with indirect jump GPIs, resulting in the elimination of call-ending gadgets and the introduction of jump-ending gadgets. Return-ending gadgets will also be eliminated if the indirect jump replaces the return instruction completely, although this only occurs if the tail call occurs on all paths to the return instruction. \verb|Clang|'s code generation convention for function epilogues includes several \verb|pop| instructions that restore the values of callee-saved registers (i.e., \verb|RBX|, \verb|RBP|, \verb|R12-R15|) before returning to the calling function. When the function-terminating return instruction preceded by these \verb|pop| instructions is replaced by an indirect jump GPI, JOP data loader gadgets are produced. 

\subsubsection{Concrete Example} 
\begin{figure}[ht]
  \centering
  \includegraphics[width=0.7\linewidth]{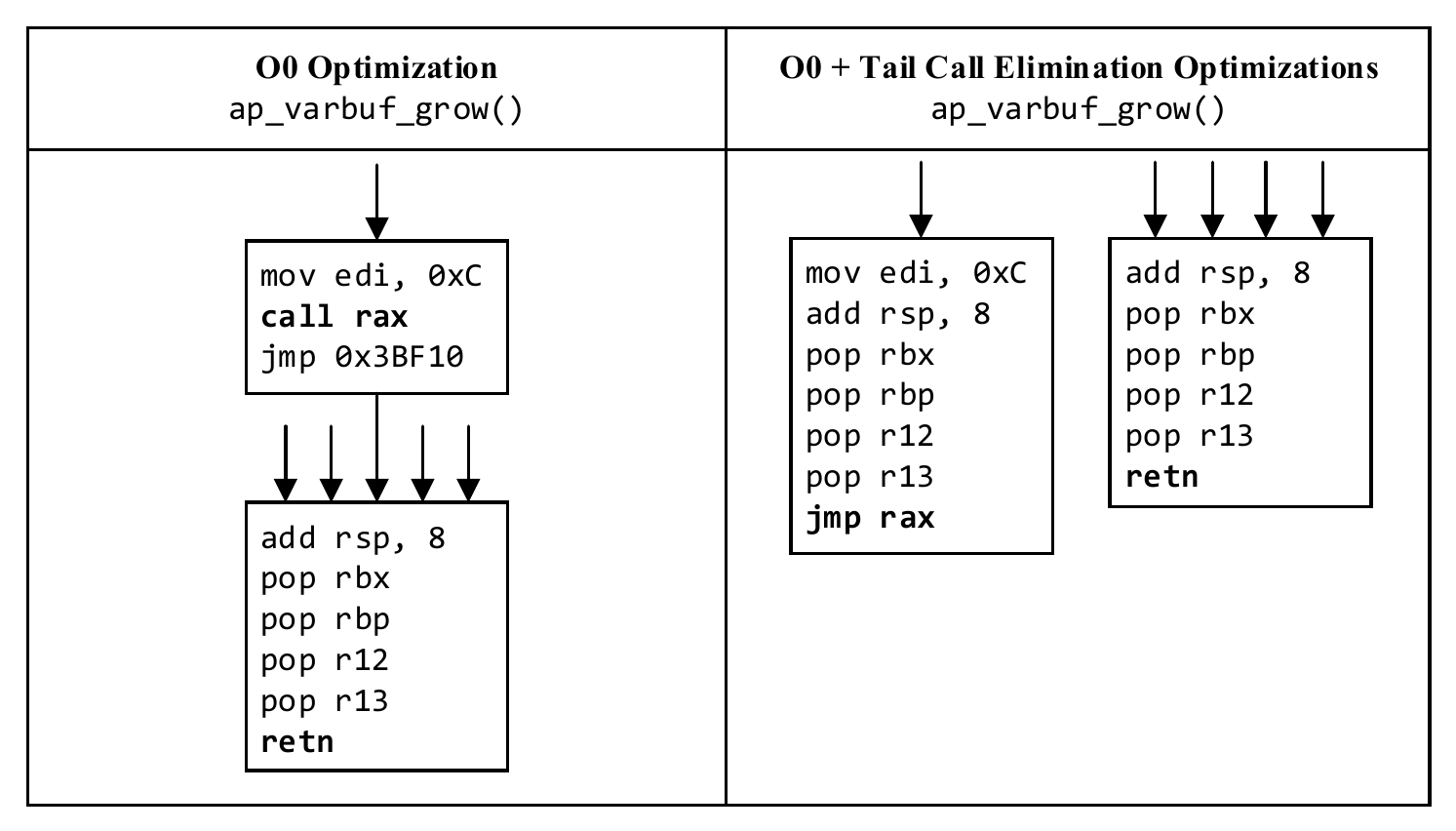}
  \caption{Special Purpose Gadget Introduction caused by Clang's Tail Call Elimination Optimization}
  \label{fig:httpd-tail-call-elim}
\end{figure}

An example of this behavior in \verb|httpd| is shown in Figure~\ref{fig:httpd-tail-call-elim}. Prior to optimization, the code snippet contains two GPIs depicted in \textbf{bold} print. After TCE the first GPI, \verb|call rax|, is transformed into a new GPI, \verb|jmp rax|, which produces a JOP data loader gadget as well as functional JOP gadgets. Minimal changes are observed with respect to the other GPI, \verb|retn|.

\subsubsection{Mitigation Strategy} In this case, the sequence of instructions representing the JOP data loader gadget is intentionally placed by the compiler during  optimization and cannot be avoided without forgoing its benefits. Disabling TCE during compilation is a straightforward mitigation strategy that has become common practice for other optimizations that introduce security weaknesses (e.g., dead store elimination)  \cite{yang2017dead}. We evaluate the performance impacts of disabling TCE in Section \ref{section:selective_disable}. Since this behavior only occurs with indirect tail calls, \verb|Clang|'s TCE pass could also be patched to optimize direct tail calls only. In either case, avoiding JOP data loader gadget introduction may not be worth forgoing TCE's performance benefits and is ultimately subject to risk-reward considerations that vary depending on the circumstances. 

\section{Mitigation Strategy Implementation and Evaluation}
\label{section:mitigation}

To address our sixth motivating question \{\emph{What is the performance cost of implementing such mitigations?}\}, we implemented and evaluated our proposed mitigation strategies. To mitigate GPI duplication and transformation-induced gadget introduction, we implemented five post-production binary recompiler passes that merge duplicated GPIs and eliminate layout-based unintended gadgets. To address intentional introduction of special purpose gadgets in \verb|Clang|'s TCE optimization, we measure the performance impact of disabling it during compilation.

\subsection{Post-Production Recompilation to Merge GPIs and Eliminate Unintended Gadgets}
\label{section:postprod_merging}

The purpose of our binary recompiler passes is to mitigate the undesirable optimization behaviors we observed without significantly impacting run-time performance or code size. We implemented two mitigation passes that address GPI duplication behaviors: one that merges \verb|retn| GPIs and one that merges indirect \verb|jmp| GPIs. We also implemented three mitigation passes that address layout-based gadgets introduced indirectly by optimizations: one that eliminates GPI encodings across instruction boundaries, one that eliminates GPI encodings in direct jump instruction displacements via short \verb|nop| sleds, and one that eliminates GPI encodings in direct call displacements via function re-ordering. 

As these two root cause behaviors are common to several different optimizations and compilers, our implementations target undesirable behaviors, not specific optimization algorithms. It is worth noting that these mitigation passes do not differentiate between GPIs introduced via optimization versus instances where the GPI is placed by the compiler during code generation. Since we expect the performance costs of mitigation to be negligible, we intentionally designed these mitigation passes to aggressively eliminate GPIs to achieve maximum benefit. To support further research in this important area, we have made the source code for our mitigation passes publicly available (see Section \ref{app:artifacts}).

\subsubsection{Implementation of GPI Merging Passes}
We observe that the benefits of duplicating GPIs in the manner depicted in Figure \ref{fig:httpd-omit-frame-ptr} are small, saving between one and five bytes of code size per eliminated jump as well as the run time required to execute the jump. Since these benefits are secondary to the original optimization, it is possible to merge all occurrences of a particular GPI within a function to a single instance and retain the optimization's primary benefits. Our return merging algorithm operates by scanning each function to find all of its return-type instructions. If multiple return-type instructions are found, one is arbitrarily selected to be retained. All other return-type instructions are replaced with an unconditional jump to the retained instruction, effectively merging multiple return-type instructions into a single instance. Our indirect jump merging algorithm operates in a manner similar to our return merging algorithm. The primary difference is that indirect jumps can only be merged if they target the same register. During the function scan, indirect jump instructions found are placed into a map data structure in which each register is mapped to a list of indirect jump instructions targeting it. When the scan is complete, any indirect jump instructions targeting the same register are merged in the same manner as the return merging algorithm.

Both of our GPI merging passes preserve program semantics. The unconditional jump instructions inserted in place of eliminated GPIs do not alter program state other than to transfer control to the retained GPI. In both of our passes, GPIs replaced by an unconditional jump are identical to the retained GPI targeted by the inserted jump, resulting in no change to program semantics.

\subsubsection{Implementation of Layout-Based Gadget Elimination Passes} In the case of layout-based GPIs introduced by transformations, such as the instance depicted in Figure \ref{fig:libcurl-rop}, we observe that small non-semantic changes to the program's layout are sufficient to eliminate these GPIs. For problematic layout encodings that are intraprocedural (i.e., instruction or basic block level), insertion of short 1 or 2 byte \verb|nop| sleds can shift problematic encodings to benign ones with a negligible performance impact. We implemented two such transformations: instruction boundary widening and jump displacement sledding. For problematic layouts that are interprocedural (i.e., direct function calls) \verb|nop| sleds must be much larger than 1-2 bytes in general to shift the displacement to a benign encoding due to function sizes, inter-function padding, etc. As such, this approach is not ideal as it swells code size and can have negative impacts on run-time performance. Instead, we implemented a program-wide transformation that reorders functions in the program layout in order to shift call displacements to benign encodings. This approach obviates the need for large \verb|nop| sleds and thus avoids their negative performance impacts, though it must be employed carefully to avoid causing slowdowns due to compromised cache locality.

Our instruction boundary widening algorithm operates by scanning each pair of contiguous instructions in the program layout to identify instances where multi-byte GPIs are encoded by the last byte of the first instruction and the first byte of the second instruction. For example, the instruction sequence \verb|{je 0xffcf; or PTR [rbx+0x4],0x2;}| is encoded as \verb|{74 CD; 80 4B 04 02;}|. The byte sequence \verb|CD 80| is contained within this sequence and encodes the syscall GPI \verb|int 0x80|. When found, our algorithm inserts a single \verb|nop| instruction to widen the boundary between the instructions (i.e., \verb|{je 0xffcf; nop; or PTR [rbx+0x4],0x2}|), eliminating the GPI encoding.

Our jump displacement sledding algorithm first scans each function to identify problematic near or short jump displacement encodings such as the one depicted in Figure \ref{fig:libcurl-rop}. For each function with problematic encodings, one encoding is selected at random to be shifted via insertion of a short (i.e., 1 or 2 bytes depending on the encoding) \verb|nop| sled. Sleds are inserted before backward jumps and before a forward jump's target to shift the problematic displacement to a benign encoding. Since adjusting the function layout in this manner can affect other jump displacements within the function, this process (i.e., scan, randomly shift an encoding per function) must be conducted iteratively until each function no longer contains problematic displacements. In certain rare cases, the layout of two or more jumps may be interrelated such that correcting a problematic encoding for one jump causes an interrelated jump encoding to become problematic, and vice versa. To ensure our algorithm terminates and does not insert excessively long and ineffective sleds, our algorithm is capable of detecting repeated failures to reduce the number of problematic encodings for a given function. If detected, our algorithm excludes the function from further iterations.

Our function reordering algorithm operates by first scanning the program to identify problematic call displacement encodings, the functions they target, and the shift size in bytes required to make the encoding benign. The algorithm then chooses a function with problematic encodings at random and determines the largest shift size required to address all of that function's problematic encodings. Then, the function is moved forward or backward in the program layout by swapping its position with a neighboring function until it has moved the minimum number of bytes necessary to shift the encodings. By moving functions by the minimum distance possible, the algorithm avoids large changes to code locality. As was the case with jump displacement sledding, scanning and re-ordering must be performed iteratively since each adjustment to program layout potentially fixes or creates other problematic encodings. In certain rare cases, the layout of two or more functions may be interrelated such that correcting a problematic encoding for one call causes an interrelated call encoding to become problematic, and vice versa. To ensure termination, this algorithm detects repeated failures to reduce the global number of problematic encodings. If a sufficient number of consecutive failures is detected, our algorithm terminates having arrived at a local minimum.

All three of our layout-based gadget elimination passes preserve program semantics. In the case of instruction boundary widening and jump displacement sledding, our mitigations only alter the binary by inserting \verb|nop| instructions, which by definition do not alter program behavior. Additionally, our function reordering algorithm alters only the relative positioning of functions in the binary layout and does not insert, modify, or remove instructions. Since our passes are only compatible with position-independent code, there is no possibility of breaking position-dependent control-flow transfers when reordering functions.

\subsubsection{Egalito Binary Recompiler.} We developed our mitigation passes for Egalito~\cite{williams2020egalito}, a binary recompiler that lifts position-independent ELF binaries to a layout-agnostic intermediate representation (EIR) that supports  arbitrary transformations and recompilation back to an ELF binary. Lifting to EIR requires precise disassembly of the binary, which Egalito accomplishes by using metadata present in the position-independent code (PIC) for analysis. PIC is dominant in most Linux binaries, and allows for full disassembly in the vast majority of cases. 

We selected Egalito as the engine for our mitigation passes for four reasons. First, it supports \textbf{machine-dependent} transformations that can readily address CRA gadgets, whereas existing compiler toolchains primarily support machine-independent optimization (e.g., LLVM~\cite{10.5555/977395.977673}). Second, Egalito is compiler agnostic; a single implementation of our mitigation passes can be used on binaries regardless of the compiler used to produce them. Third, implementing our passes in Egalito allows our work to address legacy binaries, provided that they are PIC. Finally, Egalito provides a Union ELF mode in which the input binary and its linked libraries are combined into a single output ELF prior during re-compilation. This feature allows our mitigation passes to be readily applied to library code in a single operation and also removes unnecessary library code and their constituent gadgets as an added benefit. While not employed or evaluated in this paper, we identify exploring the benefits of Egalito's Union ELF mode as promising future work.

\subsubsection{Evaluation of Gadget Set Impacts} To determine how effectively our mitigation passes reduce the availability and utility of CRA gadgets, we used Egalito to apply them to \verb|GCC| and \verb|Clang| O3 variants of our benchmark programs.\footnote{Several O3 program variants in our study were originally built as position-dependent code. They were rebuilt as PIC for this evaluation and thus metric values reported in this section may differ from those in Section~\ref{section:study}.} We then used GSA to measure the change in gadget set metrics after recompilation with our passes. Three of our benchmark programs were excluded from this evaluation because they were not compatible with Egalito (see Section \ref{sec:correctness}). Our results are reported in Table~\ref{tab:mitigation_O3}.

\begin{table*}
    \footnotesize
    \caption{Effects of Mitigation Passes on Gadget Set Metrics for O3 Variants}
    \label{tab:mitigation_O3}
    {\rowcolors{4}{lightgray!50}{white}
\begin{tabular}{|c|l|c|c|c|c|c|c|}
    \cline{3-8}
     \multicolumn{2}{c|}{}  & \multicolumn{2}{c|}{\makecell{\textbf{Functional Gadget} \\ \textbf{Set Expressivity}}} & \multicolumn{2}{c|}{\makecell{\textbf{Functional Gadget} \\ \textbf{Set Quality}}} &  \multicolumn{2}{c|}{\makecell{\textbf{S.P. Gadget} \\ \textbf{Availability}}} \\
     \hline
   \multicolumn{2}{|c|}{ \textbf{Benchmark}} & \textbf{O3} & $\Delta$\textbf{MP} & \textbf{O3} & $\Delta$\textbf{MP} &  \textbf{O3} & $\Delta$\textbf{MP}  \\
    \hline
    & Bftpd & 8/21/8 &  (0/-3/1) & 440 / 1.7 & (-118 / 0.1) & 4 & (-2) \\
    \cellcolor{white} & libcUrl & 9/33/16 & (0/-2/-1) & 6866 / 1.7  & (-2009 / 0.1) & 7 & (-2) \\
    & git & 11/35/17 & (-2/-1/0) & 13017 / 1.7  & (-4773 / 0) & 7 & (0) \\
    \cellcolor{white}& gzip & 6/26/12 & (2/3/1) & 575 / 1.6  & (-141 / 0) & 5 & (-1) \\
    & httpd & 9/33/17 & (0/-1/-2) & 5301 / 1.7  & (-2717 / 0) & 7 & (-3) \\
    \cellcolor{white}& libsqlite & 11/35/15 & (-3/-2/0) & 6512 / 1.7  & (-2194 / 0.1) & 7 & (-1) \\
    & 401.bzip2 & 8/25/10  & (0/2/-1) & 403 / 1.6 & (-101 / -0.1) & 2 & (-1) \\
    \cellcolor{white}& 403.gcc & 11/35/17  & (-1/-1/0) & 21636 / 1.6  & (-10277 / 0.1) & 7 & (0) \\
    & 429.mcf & 6/20/6  & (-1/-1/0) & 129 / 1.8 &  (-21 / 0.1) & 1 & (0) \\
    \cellcolor{white}& 433.milc & 8/28/12  & (-1/-1/0) & 891 / 1.7  & (-272 / -0.1) & 5 & (-1) \\
    & 444.namd & 8/30/12  & (0/-6/-4) & 512 / 1.6  & (-112 / 0.2) & 6 & (-5) \\
    \cellcolor{white}& 445.gobmk & 11/35/17  & (-2/-2/-2) & 5262 / 1.8 &  (-2706 / -0.2) & 7 & (-3) \\
    & 456.hmmer & 9/32/15  & (0/0/-2) & 2335 / 1.6 &  (-1125 / 0.1) & 7 & (-4) \\
    \cellcolor{white}& 458.sjeng & 10/29/13  & (-2/-1/0) & 863 / 1.5 &  (-289 / 0.2) & 4 & (-3) \\
    & 462.libquantum & 7/25/9  & (0/-2/-3) & 323 / 1.4 & (-75 / 0.2) & 2 &  (-1) \\
    \cellcolor{white}& 470.lbm & 6/20/8 & (0/0/-2) & 107 / 1.5 & (-15 / 0.1) & 1 & (0) \\
    \multirow{-18}{*}{\rotatebox[origin=c]{90}{\textbf{GCC Variants}}} & 
    482.sphinx3 & 9/31/13 & (0/0/0) & 1289 / 1.6  & (-474 / 0) & 5 & (-4) \\
    \hline
    \cellcolor{white}& Bftpd & 8/21/8  & (0/-3/1) & 435 / 1.7 &  (-112 / 0.1) & 3 & (-1) \\
    & libcUrl & 9/32/16  & (0/0/-1) & 6750 / 1.7 & (-1511 / 0) & 6 & (0) \\
    \cellcolor{white}& git & 11/35/17  & (-2/-1/0) & 13846 / 1.7 & (-5234 / 0) & 7 & (-1) \\
    & gzip & 7/26/8  & (-1/-6/1) & 417 / 1.6 &  (-92 / 0) & 4 & (-2) \\
    \cellcolor{white}& httpd & 9/34/17 &  (0/-2/-1) & 4500 / 1.6 &  (-1992 / 0) & 7 & (-3) \\
    & libsqlite & 11/34/16 & (-2/-1/0) & 6893 / 1.6  & (-2179 / 0.1) & 7 & (-2) \\
    \cellcolor{white}& 401.bzip2 & 7/26/9 & (-1/-2/-1) & 334 / 1.5  & (-53 / 0.1) & 2 & (-1) \\
    & 403.gcc & 11/35/17 & (-2/-1/0) & 18910 / 1.6   & (-8205 / 0.2) & 7 & (0) \\
    \cellcolor{white}& 429.mcf & 6/15/6  & (-1/-1/0) & 140 / 1.8 &  (-23 / -0.1) & 0 & (0) \\
    & 433.milc & 8/30/12  & (-1/-3/-2) & 710 / 1.6  & (-135 / -0.1) & 2 & (0) \\
    \cellcolor{white}& 444.namd & 6/26/11 & (0/1/1) & 882 / 1.6  & (-94 / 0) & 2 & (0) \\
    & 445.gobmk & 10/34/17  & (-1/-2/-3) & 5162 / 1.8  & (-2642 / -0.1) & 7 & (-1)  \\
    \cellcolor{white}& 456.hmmer & 9/33/16  & (0/0/-2) & 2098 / 1.7  & (-1022 / 0.1) & 7 & (-4) \\
    & 458.sjeng & 10/33/15  & (0/-3/-2) & 705 / 1.8  & (-211 / -0.1) & 4 & (-4) \\
    \cellcolor{white}& 462.libquantum & 7/26/11 & (1/-1/-1) & 339 / 1.7  & (-73 / -0.2) & 1 & (-1) \\
    & 470.lbm & 5/16/4 & (0/0/0) & 112 / 1.5 &  (-24 / 0.1) & 0 & (0) \\
    \cellcolor{white}\multirow{-18}{*}{\rotatebox[origin=c]{90}{\textbf{Clang Variants}}} &
    482.sphinx3 & 8/30/13 & (0/1/1) & 1210 / 1.6 & (-370 / 0) & 4 & (-2) \\
    \hline
\end{tabular}
}

\end{table*}

Our passes were highly effective at reducing special purpose gadget diversity in recompiled binaries. They decreased the number of special purpose gadget categories available in 75\% (24 of 32) of variants with at least one type of special purpose gadget present in the baseline O3 variant. In more than half of these cases (14 of 24), multiple categories of gadgets were eliminated. Further, we observed no instances where our mitigation passes increased the number of special purpose gadget types available. On average, our mitigation passes reduced the types of special purpose gadgets available by 34\% (36.1\% for \verb|GCC| variants and 31.8\% for \verb|Clang| variants).

Our mitigation passes were also highly effective at reducing the number of useful gadgets available after recompilation. On average, our passes reduced the total number of useful gadgets by 31.8\% (33\% for \verb|GCC| variants and 30.6\% for \verb|Clang| variants). We observed a maximum reduction rate of 51\% and in no instances did our mitigation passes increase the number of useful gadgets. Additionally, we observed no change on average across all of our benchmarks to the average gadget set quality score. This indicates that our passes eliminate high-quality and low-quality gadgets at roughly the same rate.

Finally, our mitigation passes were also quite effective at reducing functional gadget set expressivity. Our passes reduced the overall expressivity of gadget sets in 78\% (28 of 34) of total variants; however we did observe an overall increase in expressivity in four variants. Across all variants and expressivity levels, our passes reduced expressivity by approximately one computational class on average. Of critical importance is our passes' performance on fully expressive variants. Of the 21 instances where a variant was fully expressive at some level, applying our passes successfully reduced 17 of these instances below the fully expressive threshold, a success rate of \textbf{81\%}.

\subsubsection{Evaluation of Performance Impacts} 
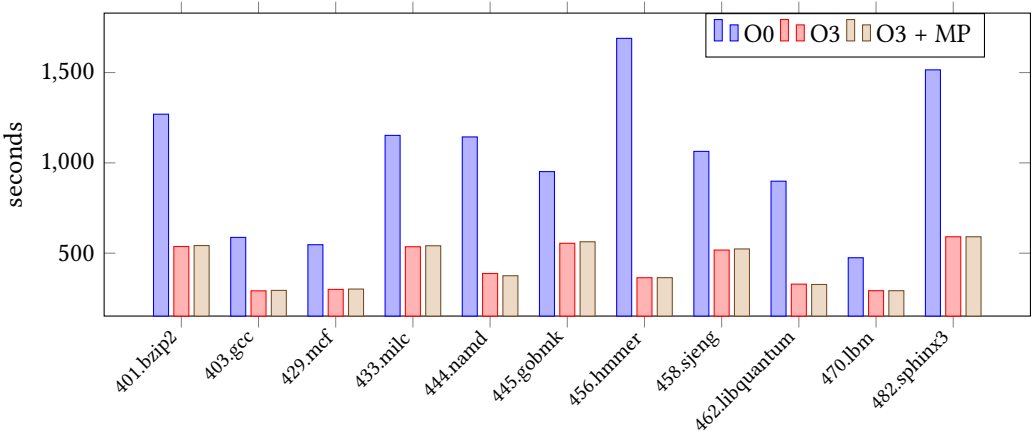
\begin{figure}
\centering
\begin{tikzpicture}
\begin{axis}[
	xtick=data,
	x tick label style={rotate=45, anchor=east, font=\footnotesize},
	symbolic x coords={401.bzip2, 403.gcc, 429.mcf, 433.milc,                    444.namd, 445.gobmk, 456.hmmer, 458.sjeng, 462.libquantum, 470.lbm, 482.sphinx3},
	ybar,
	width=\textwidth,
	height=.4\textwidth,
	bar width=.2cm,
	ylabel = seconds,
	legend style={at={(0.8,1)}, anchor=north,legend columns=-1},
]
\addplot table[x=bench,y=O0]{\clangperf};
\addplot table[x=bench,y=O3]{\clangperf};
\addplot table[x=bench,y=O3+MP]{\clangperf};
\legend{O0, O3, O3 + MP}
\end{axis}
\end{tikzpicture}
\caption{Performance Comparison of SPEC 2006 Clang Variants Recompiled with Mitigation Passes (MP)} \label{fig:clang_perf}
\end{figure}

To analyze the impact of our mitigations on execution speed, we ran the position-independent O0, O3, and recompiled variants of our SPEC 2006 benchmarks using reference workloads.\footnote{We exclude our common Linux benchmarks from this analysis due to a lack of standardized performance tests.} We recorded the total run time for each variant across three trials to determine its average performance, shown in Figures \ref{fig:gcc_perf} and \ref{fig:clang_perf}. We observed that binaries recompiled with our passes enabled saw an overall performance \textbf{improvement} of 0.2\% on average, with a maximum observed slowdown of 3.3\%. This performance improvement is not entirely unexpected, as Egalito's core recompilation process has been shown to speedup SPEC 2006 benchmarks by 1.7\% on average \cite{williams2020egalito}. We conclude that the impacts of our passes are negligible over the base performance of compiler-produced O3 variants, and are imperceptible considering the large performance improvements (i.e., 53\% speedup on average) O3 variants enjoy over O0 variants. Thus, recompiling compiler optimized code with our mitigation passes provides the best of both worlds: virtually identical run-time performance to compiler-optimized code with significantly reduced CRA gadget utility and availability.

\pgfplotstableread[row sep=\\,col sep=&]{
    bench & O0 & O3 & O3+MP  \\
    401.bzip2 & 1205.5 & 550.2 & 552.5 \\ 
    403.gcc   & 534.6  & 288.0 & 287.5 \\
    429.mcf   & 519.0  & 286.2 & 276.7 \\
    433.milc  & 889.1  & 544.5 & 544.3 \\
    444.namd  & 968.9  & 423.4 & 431.2 \\
    445.gobmk & 942.4  & 569.8 & 556.1\\
    456.hmmer & 1889.1 & 373.4 & 372.9 \\
    458.sjeng & 1046.1 & 569.5 & 582.3 \\  
    462.libquantum & 761.1  & 340.4 & 339.8 \\  
    470.lbm        & 568.2  & 340.1 & 339.8 \\
    482.sphinx3    & 1546.6 & 563.7 & 586.9 \\ 
    }\gccperf
    

\begin{figure}
\centering
\begin{tikzpicture}
\begin{axis}[
	xtick=data,
	x tick label style={rotate=45, anchor=east, font=\footnotesize},
	symbolic x coords={401.bzip2, 403.gcc, 429.mcf, 433.milc, 444.namd, 445.gobmk, 456.hmmer, 458.sjeng, 462.libquantum, 470.lbm, 482.sphinx3},
	ybar,
	ylabel = seconds,
	width=\textwidth,
	height=.4\textwidth,
	bar width=.2cm,
	legend style={at={(0.8,1)}, anchor=north,legend columns=-1},
]
\addplot table[x=bench,y=O0]{\gccperf};
\addplot table[x=bench,y=O3]{\gccperf};
\addplot table[x=bench,y=O3+MP]{\gccperf};
\legend{O0, O3, O3 + MP }
\end{axis}
\end{tikzpicture}
\caption{Performance Comparison of SPEC 2006 GCC Variants Recompiled with Mitigation Passes (MP)} \label{fig:gcc_perf}
\end{figure}
\pgfplotstableread[row sep=\\,col sep=&]{
    bench & O0 & O3 & O3+MP  \\
    401.bzip2 & 1269.6 & 536.9 & 542.3 \\
    403.gcc   & 587.4  & 291.6 & 293.9 \\ 
    429.mcf   & 546.9  & 299.5 & 301.4 \\ 
    433.milc  & 1152.2 & 535.7 & 541.1 \\ 
    444.namd  & 1143.7 & 387.6 & 375.1 \\ 
    445.gobmk & 951.7  & 554.6 & 563.2 \\ 
    456.hmmer & 1689.5 & 364.2 & 364.5 \\ 
    458.sjeng & 1063.6 & 517.3 & 523.5 \\ 
    462.libquantum & 898.6  & 328.6 & 326.7 \\ 
    470.lbm        & 474.8  & 292.4 & 292.3 \\
    482.sphinx3    & 1515.1 & 591.0 & 590.9 \\
    }\clangperf
    

To analyze the impact of our mitigation passes on code size, we compared the binary sizes of the compiler-produced O3 variants against their recompiled variants. Recompiling binaries with Egalito can result in significant changes to code size, even when no transformation passes are selected. This is due to differences in Egalito's code generation conventions regarding program layout, function padding, and control-flow. In this analysis, we are interested both in how Egalito impacts code size as well as the incremental impact of our mitigation passes. Table~\ref{tab:code_size} contains the binary sizes (in kB) of the compiler-produced position-independent O3 variant (i.e., the baseline), a control variant produced by recompiling the baseline with no passes selected, and a variant produced by recompiling the baseline with our passes enabled. In columns marked with a $\Delta$ symbol, the values in parentheses indicate the size increase caused by our mitigation passes over the control variant.

While the Egalito recompilation process can cause significant changes in code size (both positive and negative), our mitigation passes typically do not incrementally increase code size beyond that caused by Egalito. Our mitigation passes incrementally increased the binary size in roughly one-third of variants, with an average incremental increase of 6.1 kilobytes. With respect to the impact of recompilation on code size, we observed that code size impacts in our control variants were typically negligible. In 27\% (9 of 34) of variants, recompilation reduced code size, and in 53\% (18 of 34) of cases recompilation increased code size by less than 80 kilobytes. Large increases in code size were observed in the remaining seven variants, in some cases many times over the baseline size. These edge cases indicate instances where Egalito's code generation conventions can be improved.

\begin{table*}
\centering
    \footnotesize
    \caption{Impact of Mitigation Passes on Code Size (kB)}
    \label{tab:code_size}
    {\rowcolors{4}{lightgray!50}{white}
\begin{tabular}{|l|c|c|c|c|c|c|c|c|}
    \hline
    & \multicolumn{4}{c|}{\textbf{GCC}} & \multicolumn{4}{c|}{\textbf{Clang}} \\
     \cline{2-9}
   \multirow{-2}{*}{\textbf{Benchmark}}& \textbf{O3} & \textbf{Control} & \textbf{MP} & $\Delta$\textbf{MP} &  \textbf{O3} & \textbf{Control} & \textbf{MP} & $\Delta$\textbf{MP}\\
    \hline
    Bftpd & 314.6 & 116.0 & 116.0 & (0)& 99.5 & 116.0 & 116.0 & (0)\\
    libcUrl & 485.2 & 460.2 & 468.4 & (8.2) & 455.6 & 439.7 & 439.7  & (0)\\
    git & 19403.2 & 3626.3 & 3634.5 & (8.2) &  11195.0 & 3151.2 & 3155.3 & (4.1) \\
    gzip & 433.4 & 488.8 & 488.8 & (0) &  108.6 & 464.3 & 464.3 & (0)\\
    httpd & 925.6 & 910.7 & 914.8 & (4.1) &  842.3 & 837.0 & 837.0 & (0)\\
    libsqlite & 5024.5 & 927.1 & 947.6 & (20.5) & 6643.7 & 1328.5 & 1344.9 & (16.4)\\
    401.bzip2 & 110.6 & 128.3 & 132.4 & (4.1) & 98.5 & 116.1 & 120.2 & (4.1) \\
    403.gcc & 4750.1 & 5203.3 & 5313.9 & (110.6) & 4021.3 & 4507.0  & 4527.5  & (20.4) \\
    429.mcf & 26.9 & 54.5 & 54.5 & (0) &  23.0 & 58.7 & 58.7 & (0)\\
    433.milc & 197.5 & 238.8 & 238.8 & (0) & 176.9 & 222.5 & 222.5 & (0)\\
    444.namd & 349.5 & 365.9 & 365.9 & (0) & 333.1 & 353.7 & 353.7 & (0)\\
    445.gobmk & 4746.6 & 6956.4 & 6960.5 & (4.1) & 4590.0 & 6804.9 & 6804.9 & (0)\\
    456.hmmer & 410.6 & 480.5 & 480.5 & (0) & 376.5 & 452.0 & 452.0 & (0)\\
    458.sjeng & 205.1 & 2794.8 & 2794.8 & (0) & 163.8 & 2753.9 & 2753.9 & (0)\\
    462.libquantum & 55.2 & 75.0 & 75.0 & (0) & 59.7 & 79.2 & 79.2 & (0)\\
    470.lbm & 22.1 & 46.3 & 46.3 & (0) & 22.2 & 50.5 & 50.5 & (0)\\
    482.sphinx3 & 278.8 & 308.5 & 312.6 & (4.1) & 246.8 & 275.8 & 275.8 & (0)\\
    \hline
\end{tabular}
}
\end{table*}

\subsubsection{Evaluation of Correctness}
\label{sec:correctness}

To ensure that recompiling binaries with our mitigation passes enabled does not produce corrupted or otherwise invalid binaries, we performed functional testing of our recompiled variants. While the transformations performed by our mitigation passes are semantics-preserving, functional testing is necessary to ensure that errors do not occur due to synergistic effects between Egalito, our mitigation passes, the benchmark implementation, and the selected compiler.

Functional testing of our SPEC 2006 variants is straightforward as the reference workloads used in our performance evaluation also perform validity checks. For example, the reference workload for \verb|401.bzip2| compresses several test files, creating an archive file. Then, it uncompresses the archive in a temporary directory and then checks that uncompressed files are identical matches to the original test files. If they are not, the test fails there and does not report a performance result. SPEC 2006's workloads also identify when a variant under test does not terminate as expected (e.g., a segmentation fault occurs), which captures instances in which a benchmark uses language features not supported by Egalito. During functional testing of our SPEC 2006 benchmarks, we identified three such benchmarks and excluded them from our evaluation. Specifically, \verb|liblmdb| depends on a position-dependent code library and \verb|453.povray| and \verb|471.omnetpp| use C++ exceptions which are not currently supported by Egalito.

For recompiled variants of our common Linux benchmarks, we utilized developer provided functional tests to ensure their validity. For example, the larger \verb|cUrl| code repository includes a battery of tests that run against \verb|libcUrl| to verify its functionality. Only one of our benchmarks, \verb|Bftpd|, does not provide functional tests. In this case, we created a custom test suite similar to those provided for our other benchmarks to validate our recompiled \verb|Bftpd| variants.

\subsection{Disabling Clang's TCE Optimization}
\label{section:selective_disable}

\begin{figure*}[!htb]
\centering
\scalebox{0.8}{
\begin{tikzpicture}
\begin{axis}[
	xtick=data,
	x tick label style={rotate=45, anchor=east, font=\normalsize},
	symbolic x coords={401.bzip2, 429.mcf, 433.milc, 444.namd, 445.gobmk, 456.hmmer, 458.sjeng, 462.libquantum, 470.lbm, 482.sphinx3},
	ybar,
	y tick label style={font=\normalsize},
	ylabel = seconds,
	width=\textwidth,
	height=.4\textwidth,
	bar width=.4cm,
	legend style={at={(0.8,1)}, anchor=north,legend columns=-1, font=\normalsize},
]
\addplot table[x=bench,y=O3]{\clangnotce};
\addplot table[x=bench,y=O3-TCE]{\clangnotce};
\legend{O3, O3 - TCE}
\end{axis}
\end{tikzpicture}
}
\caption{Performance Comparison of SPEC 2006 Clang Variants with and without TCE Enabled} \label{fig:clang_notce}
\end{figure*}
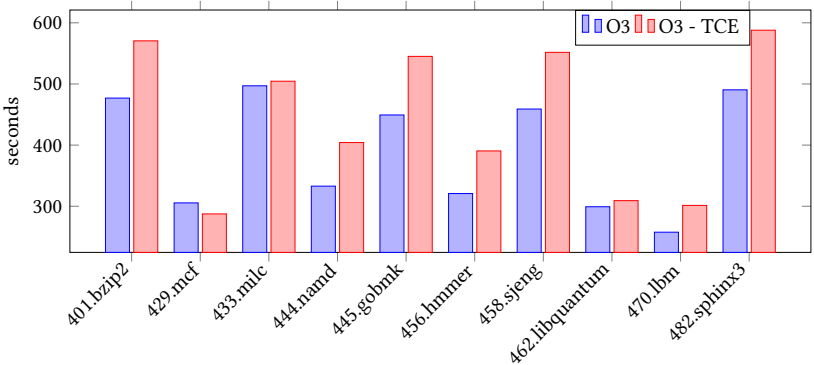

To determine the performance impact of disabling Clang's TCE optimization, we built additional Clang variants of our SPEC 2006 benchmarks at level O3 with TCE disabled. We then ran these variants using reference workloads and recorded the total runtime for each variant across three trials to determine average performance. Our results are shown in Figure \ref{fig:clang_notce}.\footnote{Note that we have excluded one benchmark, 403.gcc, from our results because Clang failed to compile it with TCE disabled.} Our results indicate that disabling TCE has a significant impact on performance, resulting in an average slowdown of 14\%. Additionally, we observed slowdowns greater than 17\% in seven of the ten variants we analyzed. Interestingly, we observed one outlier result in which disabling TCE resulted in a 5.9\% speedup, though this was the only benchmark in which a positive result was observed. 

\pgfplotstableread[row sep=\\,col sep=&]{
    bench & O3 & O3-TCE  \\
    401.bzip2 & 476.9 & 570.5 \\
    429.mcf & 305.5 & 287.5 \\
    433.milc  & 497.0 & 504.5 \\
    444.namd  & 332.9 & 404.2 \\
    445.gobmk  & 449.3 & 545.1 \\
    456.hmmer  & 320.8 & 390.5 \\
    458.sjeng  & 459.0 & 551.7 \\
    462.libquantum  & 299.2 & 309.2 \\
    470.lbm  & 257.6 & 301.4 \\
    482.sphinx3  & 490.4 & 587.9 \\
    }\clangnotce
    

Based on these results, we assess that the performance benefits of Clang's TCE optimization generally outweigh its undesirable gadget set impacts. While JOP data loader gadgets simplify exploit chain construction by providing a convenient mechanism for stack-based data injection, functional JOP exploits can be constructed without them. Thus, we conclude that the strategy of entirely disabling TCE to avoid introducing these gadgets is non-viable in most practical scenarios. A more targeted strategy of patching Clang's TCE pass to optimize direct tail-calls only may be more practical; however we consider evaluating this strategy to be outside the scope of this paper.

\section{Limitations and Considerations}

\subsection{Study Limitations} Our selection of benchmark programs is limited to C/C++ source programs compiled by \verb|GCC| and \verb|Clang| into \verb|x86-64| ELF binaries. Further research is necessary to confirm that our results generalize to build chains that utilize other programming languages (e.g., Fortran), compilers, binary formats (e.g., Windows PE), or architectures. Also, our fine-grained analysis is limited to a single optimization per variant as it was neither feasible nor necessary to consider each permutation of optimizations. However, two or more optimizations at the same level may have synergistic effects, which should be considered when applying our methodology.

Again, it is important to note that the metric calculations in our study are limited to gadgets present in our variant binaries and do not include gadgets contained in dynamically linked library code. In the context of our study, library-based gadgets are a confounding variable because shared libraries are pre-built separately from our benchmark binaries; however, these gadgets must be considered in the larger context of assessing the relative ease of crafting a code reuse attack. Additionally, defensive techniques deployed in the execution environment that prevent these gadgets from being used (e.g., position-independent code, ASLR, etc.) must also be considered when assessing gadget sets. Although our study does not include library-based gadgets in calculations, our study findings and mitigation passes are readily applicable to the build process used to create shared libraries.

\subsection{Tool Limitations} GSA's functional gadget set expressivity calculation is limited to ROP gadgets. In practice, this covers the majority of practical security use cases as JOP and COP exploits are rare in the wild. Additionally, our mitigation passes implemented with Egalito are subject to the same limitations as the tool itself; namely Egalito can only recompile position-independent code and does not support obfuscated code, inline assembly, and C++ exceptions.

\subsection{Security Considerations} When considering our findings in the context of overall software security, it is important to point out that gadget-based code reuse exploit patterns (e.g., ROP, JOP, COP) are not software vulnerabilities in and of themselves; they are powerful techniques for exploiting memory corruption vulnerabilities in the presence of  W$\oplus$X defenses. As such, the presence of expressive functional gadgets and special purpose gadgets in binaries does not necessarily make them less secure. 

However, we cannot assume or strongly prove that software is vulnerability-free in practice. As a result, it remains important to consider impacts to gadget sets when designing code generation and transformation tools. While software vulnerabilities are the point of origin for gadget-based code reuse attacks, the malicious behaviors (e.g., DoS, privilege escalation, etc.) the attacker can manifest are dependent upon the gadgets available. Thus, reducing the availability and utility of gadgets present in binaries using techniques such as those developed in this work can limit an attacker's ability to construct attacks, which in turn positively impacts overall software security. 

This work endeavors to show that compiler (and recompiler) techniques can produce performant binaries with gadget sets that are minimally useful to an attacker in the event that the program contains an unknown memory corruption vulnerability. While this approach does not eliminate software vulnerabilities, it does achieve important security objectives such as cyber hygiene and defense-in-depth.
\section{Related Work}

The mitigation strategies we have proposed and evaluated in Section \ref{section:postprod_merging} can be classified as gadget-based CRA defenses similar in nature to those described in Section \ref{section:defenses}, although our approach differs significantly from prior work. In contrast to binary retrofitting transformations that add costly dynamic (i.e., run-time) control-flow protections to binaries, our approach is fully static as to avoid run-time penalties. Despite this key difference, we consider our work to be complementary to these approaches; our proposed mitigation passes can potentially reduce the overhead costs of employing dynamic defenses by reducing the number of distinct control-flow transfers that require instrumentation and run-time protection.

With respect to compiler-based defenses, our work is similar in several ways to G-Free~\cite{onarlioglu}, an assembler injection-based solution for \verb|GCC| that eliminates unintended gadgets via transformation and inserts run-time protections for compiler-placed GPIs. Two of our transformation techniques for eliminating unintended gadgets, instruction barrier widening and jump displacement sledding, operate similarly to G-Free's. However, our approach differs in many ways. First, our approach does not require source code and is narrower in scope. Our mitigation passes are designed to address undesirable optimization behaviors identified in our study, whereas G-Free addresses other sources of unintended gadgets such as problematic register allocations during code generation. Second, our approach avoids costly run-time overheads by eliminating intended GPIs where possible via merging passes rather than inserting run-time protections. Our approach also avoids costly code size overheads by exercising direct control over program layout with Egalito. G-Free's design lacks this control and may require very large numbers of \verb|nop| instructions to eliminate unintended gadgets in call displacements and offsets. Once again, we view our approach to be complementary in that it may be used to perform gadget elimination in a manner that reduces G-Free's relatively high overhead costs (average 3.1\% slowdown and 25.9\% increase in code size).

\section{Artifact Availability}
\label{app:artifacts}
\begin{enumerate}
    \item \textbf{Study Data.} Our corpus of program variants and their associated GSA output is available at: \url{https://github.com/michaelbrownuc/compiler-opt-gadget-dataset}
    \item \textbf{Recompiler Passes.} The source code for our Egalito recompiler passes is available at: \url{https://github.com/michaelbrownuc/egalito-gadgets}
    \item \textbf{Virtual Machine.} A virtual machine in \verb|ova| format containing the above artifacts as well as other supplementary material and tools is available at: \url{https://doi.org/10.5281/zenodo.5424844} \cite{michael_d_brown_2021_5424844}
\end{enumerate}
\section{Conclusion}

In this work, we developed a data-driven methodology for studying of the impact of compiler optimizations on code reuse gadget sets. Employing this methodology, our coarse- and fine-grained analysis of optimization behaviors revealed that sets of code reuse gadgets in optimized binaries are significantly more useful for constructing CRA exploit chains than those in unoptimized binaries. We identified the root causes of our observations through differential binary analysis, and proposed potential mitigation strategies for them. We demonstrated that implementing our mitigation strategies for GPI duplication and transformation-induced gadget introduction as post-production recompiler passes can significantly reduce the availability and utility of code reuse gadgets in optimized code. Further, we demonstrated that these benefits can be obtained with negligible performance impact through a performance analysis of binaries transformed with our recompiler passes. Finally, we evaluated the performance costs of disabling \verb|Clang|'s TCE optimization and determined that this strategy is likely to be too costly in practical scenarios.

\begin{acks}
We would like to thank all of our reviewers for their helpful feedback. We also thank David Williams-King for making Egalito available for this research and for his assistance in resolving technical issues related to our work. Finally, we would also like to thank Tejas Vedantham and Chris Porter for their assistance with generating portions of the experimental data used in this study.
\end{acks}

\bibliographystyle{ACM-Reference-Format}
\bibliography{paper.bib}



\end{document}